\documentclass[a4paper,12pt]{article}%
\usepackage{amsmath}%
\usepackage{txfonts}%
\usepackage{amsfonts}%
\usepackage{amssymb}%
\usepackage{graphicx}
\usepackage{wrapfig}
\usepackage{color}
\usepackage{setspace}
\usepackage{esint}
\usepackage[rightcaption]{sidecap}
\usepackage{wrapfig}
\usepackage{lipsum}
\usepackage{caption}
\usepackage{longtable}
\usepackage[T2A,T1]{fontenc}
\usepackage[utf8]{inputenc}
\usepackage[russian,english]{babel}

\usepackage{etoolbox}
\makeatletter
\patchcmd{\@maketitle}{\begin{center}}{\begin{flushleft}}{}{}
\patchcmd{\@maketitle}{\begin{tabular}[t]{c}}{\begin{tabular}[t]{@{}l}}{}{}
\patchcmd{\@maketitle}{\end{center}}{\end{flushleft}}{}{}
\makeatother

\numberwithin{equation}{section}

\newcommand{\be}{\begin{equation}}
\newcommand{\ee}{\end{equation}}

\usepackage{anysize}
\marginsize{2cm}{2cm}{2cm}{2cm}
\usepackage{mathptmx}
-1in\relax 
\begin{document}

\begin{center}
\LARGE
\textbf{Guided Self Replicating Factory for Colonization of Solar System}
\normalsize
\vskip1cm
\begin{tabular}{ll}
\Large
\Large Mikhail V. Shubov         \\
\normalsize
University of MA Lowell          \\
One University Ave,              \\
Lowell, MA 01854                 \\
E-mail: viktor\_shubov@uml.edu   \\
\end{tabular}
\end{center}

\begin{center}
\textbf{Abstract}
\end{center}
\begin{quote}\begin{quote}
In this work, we present the concept of guided self-replicating factory (GSFR).  This factory would be established as a colony on the Moon, Mercury, Mars, or Asteroid Belt.  GSFR would grow by using in situ materials in order to manufacture machines.  The time it takes for GSFR to double it's mass and electric power output is called a doubling period.  After about 50 doubling periods, GSFR would become a Dyson Sphere civilization, which can house $10^{16}$ people, $10^{19}\ tons$ of structure and machinery and have a gross electric power of $3 \cdot 10^{24}\ W$.
GSRF would contain four parts.
The electric power station would generate electricity for all other parts.  Material production system would gather ores or native metals and produce metals, plastic, and propellant.
Material shaping and assembly system would shape metal and plastic into details. These details would be assembled into machines.  These machines would be added to all four parts of GSFR.
The space transportation system would use fuel manufactured in situ manufactured in situ in order to facilitate transportation of machinery and astronauts from Lower Earth Orbit (LEO) to the colony.
During later stages, the space transportation system would also use spaceships fully or partially manufactured from space colony material.
\end{quote}\end{quote}

\section{Introduction}
Space Colonization holds the promise of opening almost unlimited resources and ushering the New Era for Human Civilization.  The resources contained within the Solar System can support a civilization much more populous and advanced than the modern one.  A civilization encompassing the whole Solar System would be able to support a population of 10 quadrillion $\big( 10^{16}\big)$ people at material living standards vastly superior to those in USA 2020 \cite{skymine}.  Colonization of the Solar System will be an extraordinary important step for Humankind.

The first resource widely available in the Solar System is energy.  It is the most essential resource for modern industry and civilization \cite{kard, E1}.  The solar energy in space is almost limitless -- the Sun's thermal power is $3.86 \cdot 10^{26}\ W$ \cite[p.14-2]{crc}.  A future civilization, which would harvest 20\% of that power with 25\% efficiency, will have energy production of $1.7 \cdot 10^{26}\ kWh/year$.  The current global energy production in all forms is $7 \cdot 10^{13}\ kWh/year$ \cite{wes} -- 2.4 trillion times less.
The scheme for harvesting this energy is called the Dyson Sphere \cite{Dyson}.  The Dyson Sphere would consist of a multitude of space habitats orbiting the Sun.  Each habitat would be equipped with photovoltaic or solar thermal electric power station.

The second resource widely available in the Solar System is material.  The Asteroid Belt contains about 3 $Exatons$ of material composed of metal silicates, carbon compounds, water, and pure metals \cite{ABM}.  An \textbf{Exaton}, is a unit of mass equivalent to $10^{18}$ $tons$.  Most asteroids are of a carbonaceous type \cite{Asteroids}.  Carbon is very useful for production of food for space travelers, production of fuel for propulsion within space, and for production of plastics for space habitat structures.  High quality steel is also an abundant resource in space, e.g., asteroid 16 Psyche contains $10^{16}\ tons$ of nickel-rich steel \cite{Psycho}.  Additional material for comfortable habitats can be obtained from Mercury, satellites of gas giant planets, and Kuiper Belt objects \cite{Kuiper}.  Kuiper Belt contains about 120 $Exatons$ of material -- mainly water, ammonia, and carbon compounds \cite{MKuiper}.  Planet Mercury contains 330 Exatons of material composed of metal silicates, carbon compounds, and pure metals \cite[p.14-2]{crc}.  Satellites of Jupiter and Saturn contain at least 10 Exatons of water and hydrocarbons \cite[p.14-4]{crc}.  Given the data above, it is possible to construct a total habitat space of 100 Exatons.

This is enough to provide 10,000 tons of structure for each of 10 quadrillion inhabitants of the Solar System.  The mass of modern luxury cruise liners can be approximated by multiplying 85 $tons$ by the number of cabins \cite{CruiseShips}.  Space habitats will have about 120 times more structural material per inhabitant, and habitat material will be more advanced.  This will provide a material standard of living suitable for Solar System Civilization.

The colonization of the Solar System was originally proposed by Konstantin Tsiolkovsky in 1903 \cite{Tsialkovski}.   Technology necessary for space colonization appeared only in the early 1970s.  During the 1970s, many detailed studies and projects for space colonization have been published \cite{habitats,1974}.  Due to very high initial costs, the projects of space colonization developed in the 1970s and 1980s did not become a reality.  Humankind may have to wait many decades before space colonization becomes a reality.  But even if the projects, which have been planned for 1980s, will only be realized during the 2080s, it is still a great step forward for Humankind.

In the author's opinion, the colonization of the Solar System will begin with partial colonization of the Moon and colonization of Mercury.  The Lunar outpost will produce hydrogen propellant for spaceships going to Mercury.  The colony on Mercury will launch material into Mercury's orbit and construct Space Trams.  A Space Tram is a space transportation equipped with an electric power station and an electric thruster.  Space Trams would open the gates toward colonization of the Solar System.

The colonization of the Solar System is assumed to contain at least four stages.  During the First Stage, the Moon, Mercury, Mars, the Asteroid Belt, and possibly outer planet satellites will be thoroughly explored by astronauts as well as robotic probes.  On Earth, further stages of colonization will be engineered in great detail.

During the Second Stage, a one or more colony would be set up and technology for utilization of in situ resources will be delivered there.  We expect that the first such colony is set up on the Moon.  Later colonies may be set up on Mercury, the  Asteroid Belt, and/or on Mars.  Electric power generation, ore refining and smelting, chemical, and metallurgical plants will be set up.  Material produced in these plants will enable the colony to expand existing plants and to built new ones.  The new plants will still need sophisticated equipment delivered from Earth.  The first two stages of Solar System colonization will be costly \cite{cost}, but they should soon turn into the next one.

During the Third Stage, almost all machines will be manufactured from materials available in situ, while some high technology will still be delivered from the Earth \cite{planz01}. During that stage, one or more colonies around the Solar System will start manufacturing ships for transportation around the Solar System.  Third stage will be marked by exponential growth of colonies.

During the Fourth Stage, almost all machinery will be manufactured on colonies around the Solar System.  Exponential growth will continue.  Colonies throughout the Solar System will be ready to accept numerous inhabitants.  Vehicles for transportation around the Solar System will be manufactured in large numbers.  They will open the gates to colonization of the Outer Solar System.
At this point, we are unable to foresee all the further stages in the growth of space colonies.

\section{Guided Self Replicating Factory (GSRF) concept}
\subsection{Self replicating factories (SRF)}
The concept of \textbf{self-replicating factory} (SRF) to be used for Space Colonization has been studied since 1940s \cite{SRM02,SRM1,SRM2}.  Some works on self replicating machines are purely or mostly theoretical \cite{SRM3,SRM4}.
Self-replicating lunar factories have been designed by university professors \cite{SRM8,SRM9}.
Advanced projects consider self-replicators on nanoscale \cite{SRM03}.

No SRF has been built so far.  Such a system may not be feasible for a long time.  At this point, we have no way of knowing how long it will take to construct SRF.  It is unknown how long it will take to surmount known difficulties.  Unforeseen difficulties may arise.

Many works of Science Fiction written between XVIIth and XIXth centuries contain concepts which are technically feasible, but have been implemented yet \cite{SciFi1,SciFi3}.  Travel to the Moon by a multi-stage rocket has first been described by Cyrano de Bergerac in 1655 \cite{SciFi1}.
Civilization based on Solar Power and fully mechanized production was envisioned and described by John Etzler in the 1830s and 1840s \cite{Solar1831,Mechanical1841}.  Now we can appreciate the technical difficulties which had or still have to be surmounted in order for ideas formulated in the previous centuries to become reality.

It is possible that technology needed to build an SRF are far more advanced than anything we can imagine in 2021.  Possibly, an SRF will be built only after the colonization of Solar System.  Such factory may need the work of trillions experts and resources of Solar System Civilization, far exceeding current capabilities.
It may also need technologies which will be developed only in the coming centuries.
Even in this case, SRF would be very useful in colonization of Milky Way Galaxy \cite{GalCol1,GalCol2}.

\subsection{Guided self replicating factories (GSRF)}

We are proposing a much simpler system, called \textbf{guided self-replicating factory} (GSRF).  This system produces electric energy.  It also produces almost all parts necessary for its growth out of in situ resources.  These resources may be regolith on the surface of the Moon or Mercury, surface resources of Mars, or resources of the Asteroid Belt.  GSRF grows exponentially.  Bootstrapping systems for space colonization have been suggested in the past \cite{Boot1,Boot2}.

Unlike SRF, GSRF is not completely autonomous.
A GSRF on Mercury, Mars, or Asteroid belt is operated by a team of astronauts.  A GSRF on the Moon is likely to be operated remotely from many computers located on Earth.
As GSRF grows, the needed team of astronauts grows along with it.  GSRF manufactures simple machines and details.  It obtains most complicated machinery from Earth.  At the later stages of growth, GSRF also produces space transports, which facilitate delivery of materials and astronauts.

The time it takes GSRF to double is called \textbf{doubling time}.  The doubling of GSRF does not mean that it produces an exact replica of itself.  \textbf{Doubling} means that GSRF doubles its mass, energy production, and machine production.

Two factors determine the usefulness of GSRF for space colonization.  First is the \textbf{seed mass} -- the minimum mass of the factory which can produce all details and machines for growth.  Designing GSRF with minimum seed mass is a complicated engineering task.  The second factor is the doubling time.  A GSRF with doubling time of many years will grow very slowly.  A GSRF with doubling time of under one year can develop into a large colony within two decades.  This colony would be a key though which the Humankind can start colonizing the Solar System.

Bootstrapping systems in which production of a resource was used to produce tools which produced more of the same resource are not new.
Discovery of Bronze and ushering of the Bronze Age enabled metalworkers to produce bronze hatchets and pickaxes.  Bronze pickaxes were used to obtain ore more efficiently than horn pickaxes.  Bronze hatchets were used to cut wood more efficiently than stone axes.  Wood was needed to smelt copper and tin ore \cite{BrAge}.
During the Industrial Revolution, coal was extracted by pneumatic hammers.  These hammers were powered by steam engines burning a small fraction of the coal produced \cite{Ugol}.

\subsection{Major components of GSRF -- overview}

GSRF contains four major components: an electric power station, a material production system, a material shaping and assembly system, and a space transportation system.
\textbf{Electric power station} produces electric energy.  A power station consisting of an array of photoelectric cells would be optimal for a lunar colony or a near-Earth spacecraft.  Another type of power station consists of a heat engine and an electric generator.  A heat engine using nuclear energy is optimal for stations on Mars and Asteroid Belt.  A heat engine using concentrated solar energy is optimal for Mercury.  Thermonuclear fusion can be an important source of energy one day, but we should not count on technology which has not been developed so far.

\textbf{Material production system} turns the regolith of Mercury, Moon, Mars, or an asteroid into useful metals, polymers, propellant, and other chemicals.
First, regolith, ores, and ice deposits are extracted by mining robots.
Second, the ores are concentrated.
Third, useful elements are extracted from the ores.
Steel, chromium, manganese, and titanium are obtained by smelting ores.
Aluminum, magnesium, calcium, sodium, and potassium are obtained by electrolysis.
Oxygen and hydrogen are obtained by water electrolysis.
Other chemicals such as methane propellant and rubber for insulation may also be produced.

\textbf{Material shaping and assembly system} fabricates details and machines out of metals and polymers.  This system is most complicated, and it has most subcomponents.  The design of such system may be a task requiring hundreds of thousands of expert-years of theoretical and experimental work.

\textbf{Space transportation system} would assist with delivery of astronauts and equipment from Earth's orbit to the Moon and different destinations within the Solar System.  Most likely, transportation from the Low Earth Orbit (LEO) to deep space would be provided by propellant manufactured on the Moon and delivered to Low Earth Orbit.  Transportation from near-Earth deep space to other destinations within the Solar System would be provided by spaceships powered by ion thrust engines.

\section{Electric Energy Generation}
In this Subsection, we describe a possible power station in the Solar System.  We also calculate its \textbf{specific mass} -- the mass of structural material and machinery necessary to produce a unit of power.
\subsection{Possible electric power plants}
\subsubsection{Thermionic Conversion}
``Thermionic energy conversion (TEC) is the direct conversion of heat into electricity by the mechanism of thermionic emission, the spontaneous ejection of hot electrons from a surface" \cite[p.1]{ThIon1}.  Electrons are emitted by a hotter plate and absorbed by a colder plate, creating an electromotive force.  Efficiencies are generally 5\% to 10\% \cite{ThIon1}.  One advantage of thermionic energy conversion is that it can use solar radiation concentrated up to 1,000 times or nuclear energy from a very high temperature heat source.  Even with low efficiency, the power output of thermionic cells can be up to 100,000 $W/m^2$ \cite[p.73]{ThIon4}.  Several experimental studies have been performed for using solar-heated thermionic electricity for space propulsion \cite[p.2237]{ThIon3}.

Photon-enhanced thermionic emission (PETE) is a new technology which may greatly enhance efficiency of very concentrated solar power if it turns out to work as expected \cite{PETE2}.  PETE may convert 1,000 times concentrated solar energy to electricity at an efficiency of 23\% \cite{PETE3}.  PETE may work very well for temperatures of 500$^o$C to 1,000$^o$C -- perfect for concentrated sunlight on Mercury \cite{PETE1}.  Silicon PETE at irradiation of 1,000 suns and 600$^o$C temperature should have efficiency of 30\% \cite{PETE7}.
The future of PETE technology is uncertain.  No working PETE converters have been produced so far.  Some researchers claim that PETE are likely to have efficiency below 8\% \cite{PETE5}.

\subsubsection{Heat engine power plants}

A motive power plant consists of a heat engine and an electric generator.  An \textbf{electric generator} converts mechanical energy into electrical energy.  Electrical generators have high efficiency -- by 1911, alternating-current generators had efficiencies of 94\% to 96\% \cite[p.43]{GEff01}.  Generators assembled from in situ material on Moon, Mercury, Mars, and Asteroid Belt should be about as efficient.  Electric generators also have high specific power.  Modern electric generators have specific power of up to 13 $kW/kg$ \cite[p.383]{Eg1}.  Most power plant generators have much lower specific power of 0.5 $kW/kg$ \cite[p.21]{Eg2}.  This specific power converts to specific mass of 2 $kg/kW$.

A \textbf{heat engine} converts thermal energy into mechanical energy.
A heat engine has three components: a heat source, a motor, and a heat sink.
%%  Heat Source
In most of the Solar System, nuclear power is the optimal heat source.  On Mercury, and possibly on the Moon, the optimal heat source is concentrated solar energy.  Concentrated solar power thermal electric stations have been designed and built on Earth \cite{CS1,CS2,CS3}.  Optimal heat source temperature for a space power plant is 800$^o$C to 900$^o$C \cite[p.5]{SPwr06}.

%% Motor
%% Potassium vapor turbine
The first type of motor is a potassium vapor turbine.  These turbines have been constructed and rigorously tested.  General Electric tested one potassium turbine for 5,000 hours.  The turbine had blade tip speed of 250 $m/s$, potassium vapor inlet temperature of 816 $^o$C, and exhaust temperature of 671 $^o$C.  Turbine blades experienced negligible erosion during the test \cite[p.5]{SPwr06}.  In order to settle on the best design, much more detailed studies are necessary.

%% Stirling engine
The second type of motor consists of Stirling engine.  A 10 $MW$ space nuclear power station using a potassium turbine should have a specific mass of 11 $kg/kWe$, while a similar power plant using a Stirling engine would have a specific mass of 21 $kg/kWe$  \cite[p.5]{SPwr05a}.  Nevertheless, Stirling engines are much simpler to manufacture than turbine engines.

%% CO2 engine
The third type of motor uses supercritical carbon dioxide as the working fluid \cite{CDioxTurb2,CDioxTurb3,CDioxTurb5}.  All the motors discussed use Brayton cycle turbine engines.  During Brayton cycle, the working fluid remains in gaseous state.  First, compressed working fluid is heated by a heat source.  Second, the working fluid expands and performs work on the turbine.  Third, expanded working fluid is cooled by the heat sink.  Forth, the cooled working fluid is compressed by the compressor.  The compressor is powered by the turbine.  The turbine generates more energy than the compressor consumes \cite{Bray}.

%% Heat sink
The heat sink is a specially constructed \textbf{radiator}.  A radiator consists of a thin metal sheet lined with thin capillary tubes which carry the cooling fluid.  Within the radiator, the fluid's thermal energy turns into black body radiation.  Radiator mass is inversely proportional to the fourth power of the temperature of the cooling fluid.  Hot cooling fluid reduces radiator mass, but it makes engine functioning more difficult.  Most designers of space power systems choose potassium as the working fluid, offering the best compromise between the engine and the radiator \cite{SPwr03,SPwr05}.

\subsection{Specific mass}

Specific masses for Deep-Space based nuclear power plants are available from previous designs.  A nuclear space power station with a potassium turbine can have specific mass of 16 $kg/kWe$ for 5 $MWe$, 13 $kg/kWe$ for 20 $MWe$, and 11 $kg/kWe$ for 100 $MWe$.  The notations $kWe$ and $MWe$ denote kilowatts and megawatts of electric power \cite[p.4]{SPwr03}.  All projected space nuclear power plants producing over 1 $MW$ electric power have specific masses below 30 $kg/kW$ \cite[p.12]{SPwr05}.
A database in \cite[p.76-78]{SPwr05} lists 83 designs for space nuclear power systems.  All multimegawatt designs have specific mass below 20 $kg/kWe$.

Estimating specific masses for solar concentrators on the Moon, Mercury, and the Asteroid Belt remains an open problem.  Nevertheless, we can estimate the specific mass of heat engines themselves.
Any motor working in a power station on the Moon, Mercury, or Asteroid Belt  is likely to use potassium vapor as the working fluid.  The first choice of motor is a turbine.  Much experimental and theoretical work has been done on the use of potassium vapor turbines for space applications \cite{KTurbo1,KTurbo2,KTurbo3}.
In a Rankine cycle, the working fluid partially condenses as it expands within the turbine.
A 1 $MW$ Rankine potassium turbine with inlet temperature of 850$^o$C has thermal efficiency of 14\% to 17\%, and a specific mass of  4 $kg/kW$ to 5 $kg/kW$ \cite[p.12]{SPwr05}.  Turbines themselves would have to be delivered from Earth.
A 1 $MW$ Stirling engine with inlet temperature of 850$^o$C has thermal efficiency of 26\% to 30\%, and a specific mass of 18 $kg/kW$ to 30 $kg/kW$ \cite[p.12]{SPwr05}.  Stirling engines with specific power of as low as 3 $kg/kW$ can be built \cite{Stirling3}.

As we mentioned above, any engine would have a radiator heat sink.  Most potassium turbines tested so far expel potassium at a temperature of 520$^o$C to 670$^o$C \cite[p.5]{SPwr06}.  At these temperatures, saturated potassium vapor has a pressure of 0.06 $atm$ to 0.4 $atm$ \cite[p.6-70]{crc}.
For turbine exhaust temperatures of 520$^o$C to 670$^o$C, the radiator has specific mass of 2.5 $kg/kW$ to 4 $kg/kW$ \cite[p.82]{SPwr06}.
A radiator using pure gas heat carrier can have heat rejection temperature of 350$^o$C and a specific mass of 1 $kg/kW$ \cite{SPwr01}.
For a 14.3\% efficient Rankine turbine engine, the power radiated away is 6 times as high as electric power produced.  Thus, the radiator specific mass defined in terms of the motive power of the engine would be 6 $kW/kg$ to 24 $kW/kg$.

\section{Manufacturing technology}
In this section, we briefly describe the passage of material from ore deposits or native metal to the finished product.  Fully designing such system would require tens of thousands of expert-years.

\subsection{Material production system}

The first step through which the ore passes is being harvested by mining robots and delivered to a mill.  As the colony expands, so does the number, size, and specialization of mining robots and delivery vehicles.

The second step is ore benefaction at the mill.  This step is applicable to colonies located on Moon, Mercury, and Mars.  On Asteroid Belt, nickel steel is available in native state. Collected ore has to be concentrated.  If particles contain more than one type of mineral, they have to be ground.  Dense ore particles can be separated from less dense particles by subjecting the mixture to centrifugal force along with shaking motion \cite[p.64]{Centrifuga}.  At the end of this step, we have concentrated ores containing iron, chromium, titanium, and manganese.  We also have concentrated ores of basic metals -- sodium, potassium, magnesium, and calcium.  We have ``tails" -- impure quartz and alumina.

The third step is ore smelting.  Energetically optimal processes for extracting metal from Lunar and Martian ores have yet to be developed.  We have some insights into extracting metal from Mercurial ores.

Mercurial regolith contains 8\% to 17\% magnesium, 5\% to 10\% aluminum, 3\% to 5\% sodium, 0.6\% to 1.7\% iron, 0.13\% to 0.15\% chromium, and 0.10\% to 0.12\% manganese \cite[p.34]{MercChComp}.  Most iron, chromium, and manganese are present in the form of sulfide ores \cite{MercC4}.  About 1\% to 3\% of mercurial regolith consists of graphite \cite{MercC1}.  About 15\% of Mercury's surface has extensive graphite deposits \cite{MercC3}.  Mercury also contains at least 9 billion tons of water ice in polar crater deposits \cite{MH2O1}.

Iron, manganese and chromium can be produced from sulfide ores by graphite reduction at high temperature:
      \be
      \label{2.01}
      \begin{split}
      FeS+Mg Si O_3+C&\to Fe+MgS+ SiO_2+CO,\\%-306\ \frac{kJ}{mol},\\
      MnS+Mg Si O_3+C&\to Mn+MgS+ SiO_2+CO,\\
      Cr_2S_3+3\ Mg Si O_3+3\ C &\to 2\ Cr+ 3\ MgS+ 3\ SiO_2+3\ CO.
      \end{split}
      \ee
Aluminum can be extracted from the ore via electrolysis using graphite anode.  Electric energy cost of rudimentary electrolysis aluminum production is 80 $MJ/kg$ \cite[p.16]{ElEff}.

\subsection{Material shaping and assembly system}

The first step of material shaping and assembly is metal machining.  During this step, raw metal such as steel and aluminum is turned into details.
We introduce the concept of \textbf{own mass production time} (OMPT) -- the time an industrial machine takes to produce its own mass in product.   Most industrial machines have OMPT of a few days.
A 25 $ton$ machine producing steel pipe 8 $mm$ in diameter and 0.5 $mm$ in thickness and working 20 hours per day has OMPT of 2 days \cite{Truba}.  Pipe produced by this machine is very useful for heat rejection systems of energy generating plants.  A steel plate rolling machine working 20 hours per day has OMPT of 2 days \cite{Tarelka}.
Nut making machines working 20 hours per day have OMPTs 10 to 26 $days$ \cite{Orechi,Orechi1}.  OMPTs are longer for lighter nuts.  Fortunately, nuts make up a small mass fraction of finished product.

A piece of raw steel, chromium, or aluminum ingot does not become final product upon passing through just one machine.  Likely, it would have to pass through about five machines and robotic assemblers.  The total OMPT for the whole process should be about 12 to 20 days.

A few uncommon details would be manufactured by 3d printing.  Metal-based \textbf{additive manufacturing} or 3d printing is ``defined as a process of joining materials to make objects from 3D model data, usually layer upon layer" \cite{AdMan}. Several types of additive manufacturing processes are used.
\begin{quote}
    \textbf{Directed Energy Deposition} utilizes thermal energy, typically from a laser, to fuse materials by melting them as they are deposited.

    \textbf{Powder Bed Fusion} uses thermal energy from a laser or electron beam to selectively fuse powder in a powder bed.

    \textbf{Sheet Lamination} uses sheets of material bonded to form a three-dimensional object.
\cite{MAM4}
\end{quote}

In Table \ref{T.08} below, we present data for several additive manufacturing machines.
\begin{center}
    \begin{tabular}{|l|l|l|l|}
    \hline
    Machine                                 & Cost        & Weight     & Printing rate          \\
    \hline
    Average for 2017 \cite[p17-20]{MAM4}     & \$1,000,000 &            & 6.3 $cm^3/h$            \\
    Sciaky’s EBAM 300 \cite{MAM7}            &             &            & 3 $kg/h$ to 9 $kg/h$    \\
    Desktop Metal System Printer \cite{MAM8} & \$49,000    & 97 $kg$    & 16 $cm^3/h$             \\
    Metalform750 \cite{MAM9}                 &             & 1,100 $kg$ & 50 $cm^3/h$ to 120 $cm^3/h$ \\
    Concept Laser Mlab Cusing \cite{MAMB}    &             & 600 $kg$   & 1 $cm^3/h$ to 5 $cm^3/h$    \\
    \hline
    \end{tabular}
    \captionof{table}{Additive Manufacturing Machines \label{T.08}}
\end{center}

It is likely that additive manufacturing will be significantly improved in the coming years.  Metal-based Additive Manufacturing is a small, but rapidly growing industry, exceeding \$1.03 billion in 2016 \cite{MAM3}.
The number of metal 3d printers sold grew nearly 80\% In 2017 \cite{MAMC}.
Additive manufacturing will likely play an important role in Solar System colonization.  Details too complicated even for additive manufacturing would be delivered from Earth.

The second step of material shaping and assembly is construction of electric stations, machines, and robots out of manufactured details.  This step is mostly done by robots.  Some parts of the process are guided by astronauts.  Most details are manufactured in situ, while some very complicated details are delivered from Earth.

\section{Space transportation}
\subsection{Space launch}
\subsubsection{Historical launch costs}
\hspace{.5cm} In the period from 1957 up to 2015, there have been 5,510 space launches, of which 5,046 were successful.  The decade with most (1,231) space launches was the 1970s  \cite{SLR}.  During the years 1990 -- 2010, 16,200 tons of payload have been launched into space and 65\% of that mass returned to the Earth with the Space Shuttle \cite{TotM}.

One of the main parameters for a launch vehicle is the cost of placing a payload into orbit.  The cost is proportional to payload mass, and it is measured in dollars per $kg$.  The evolution of launch cost is illustrated in \cite{LCost}:
\begin{center}
\includegraphics[width=12cm]{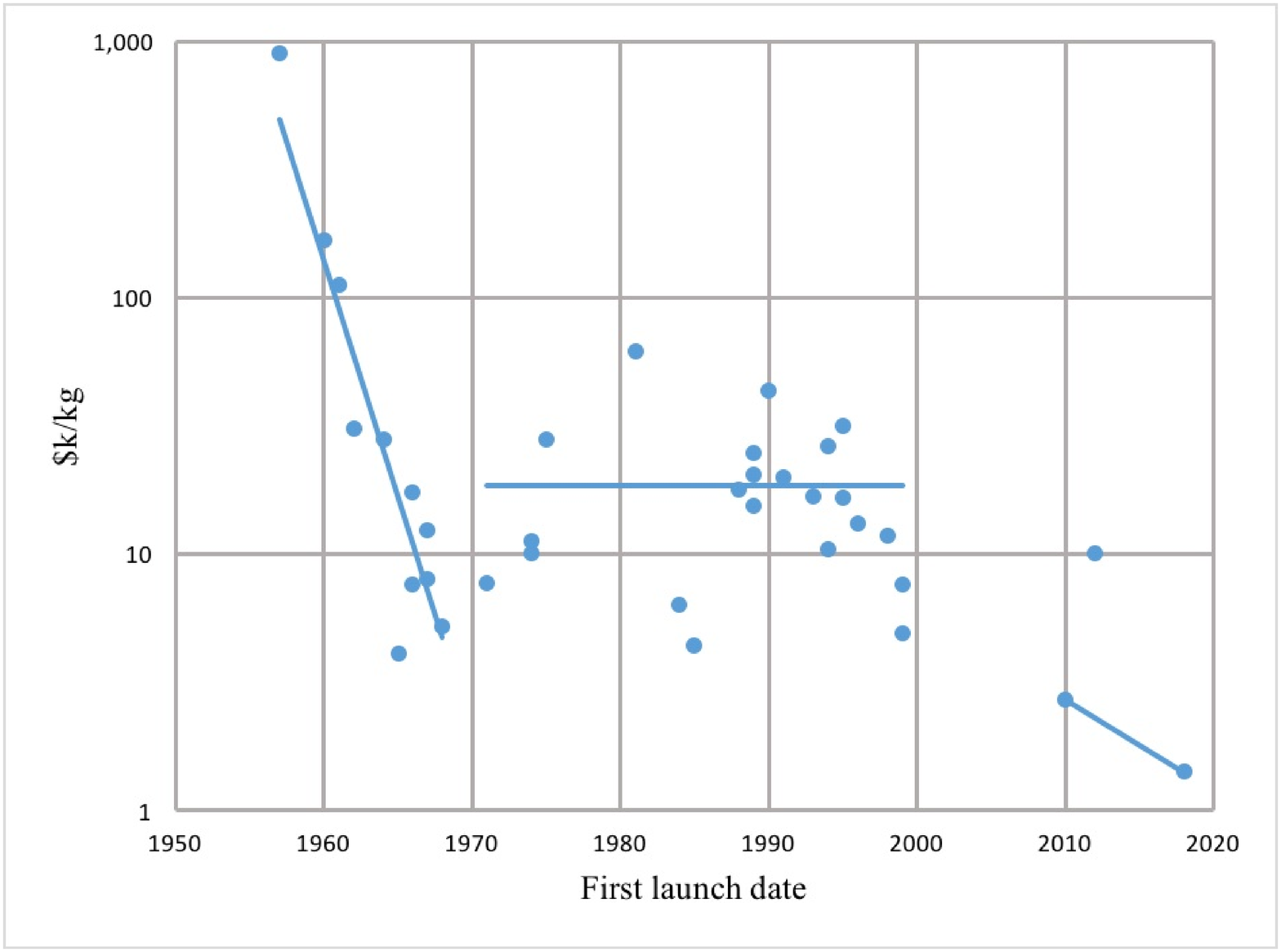}
\captionof{figure}{Launch Cost Evolution \label{F01}}
\end{center}

The cost of placing payload into the Low Earth Orbit (LEO) by the first rockets was very high.  For Vanguard launched in 1957 it was \$895,000 per $kg$ in 2018 dollars.  At first the cost dropped rapidly -- Delta E launched in 1960 had a cost of \$168,000 per $kg$ in 2018 dollars \cite[p.8]{LCost}.  Saturn V Rocket used in the Moon expedition in 1969 cost which is \$728 million in 2018 dollars \cite[p.8]{LCost}.  It could have placed 140 tons into LEO \cite[p.8]{LCost}.  Thus, by 1969, very heavy launch vehicles could place a payload into an orbit at \$5,200 per $kg$ in 2018 dollars.

In the early 1970s, most experts believed that in a few years, the cost of placing a payload into an orbit would have been \$2,200 per $kg$ in 2018 dollars \cite[p.35]{Shuttle}.  By 1980s, space launch should have cost \$400 -- \$600 per $kg$ payload in 2018 dollars \cite[p.viii]{guide}.  Unfortunately, the progress was stalled, and launch costs remained at an average of \$18,500 per $kg$ up to about 2010 \cite[p.8]{LCost}.  In 2002, the cost of LEO delivery has been \$6,900 per $kg$ for Delta4 Heavy rocket and \$18,400 per $kg$ for Titan 4B rocket \cite[p.211]{Cost1}.

In 2016, Ariane 5 ECA rocket delivered payload to LEO for \$8,500 per $kg$ \cite[p.74]{Raport01}.  Even in 2020, old and expensive launch systems are used along with the advanced SpaceX launch systems discussed in the next subsection.  It takes time for new technology to replace the old one.

\subsubsection{Launch cost reduction}
The combination of high anticipations in early 1970s and very limited progress by mid 2000s was very disappointing.  It could have taken a very long time for any progress to happen.

A giant breakthrough was accomplishes by SpaceX company.  Already by 2009, their Falcon 9 rocket delivered payload to LEO for \$2,700 per $kg$.  The next great step was the introduction of the reusable first stage.  On December 21, 2015, SpaceX made a huge step in History when the first stage of Falcon 9 spacecraft returned to the launching pad \cite[p.1]{SLVD}.  During 2016, SpaceX has successfully landed six first stage boosters \cite{Falcon2}.  By July 2019, there have been 34 successful first stage returns out of 40 attempts \cite{Falcon3}.  By 2018, SpaceX was offering LEO delivery at \$1,400 per $kg$ via Falcon Heavy \cite[p.8]{LCost}.  A two stage spacecraft with fully reusable stages will deliver payload to LEO for \$1,000 per $kg$ \cite[p.33]{Falcon1}.

Even though much has been achieved, more remains to be done.  According to the interview with SpaceX CEO Elon Musk \cite{Falcon3},
\begin{quote}
by late 2014, SpaceX determined that the mass needed for a re-entry heat shield, landing engines, and other equipment to support recovery of the second stage was at that time prohibitive, and indefinitely suspended their second-stage reusability plans for the Falcon line.
\end{quote}
As of 2020, Falcon Heavy is the least expensive launch vehicle with LEO delivery cost of \$1,400 per $kg$ \cite{FH01,FH02}.

\subsubsection{Non-rocket space launch concepts}
A \textbf{space gun} accelerates the payload in an inclined tunnel and launches it into space \cite{SpaceCannon}.  One form of a space gun is the \textbf{ram accelerator}.  It consists of a tube 4 $m$ in diameter and several $km$ long.  The tube is filled with an explosive gas combination -- like oxygen and methane.  The transport enters the tube at a velocity of 1.5 $km/s$.  The transport has scramjet  engines.  These engines accelerate the transport by burning the gas mixture.  The space gun can deliver 50,000 tons payload per year at a cost of \$2.50 per $kg$.  Neither astronauts not fragile electronics can be delivered to space by a gun due to high launch acceleration \cite[p.104-121]{NRSL1}.
At present, ram accelerator remains a purely theoretical concept.  Miniature versions of ram accelerators accelerating projectiles to high yet suborbital velocities have been tested \cite{RamAc1,RamAc2}.

A \textbf{star tram} is an electromagnetic space gun.  This star tram consists of a vacuum tube about 130 $km$ long.  The firing end of the tube is located either at a mountain top about 6 $km$ tall or is suspended in atmosphere at 10 $km$ by lifting gas balloons.  The end of the tube is inclined at 10$^o$ to horizontal.

The transport is a streamline body 2 $m$ in diameter and 13 $m$ in length.  Loaded transport has a mass of 40 $tons$.  The transport is accelerated within a vacuum tunnel by a linear induction motor.  It levitates due to magnetic field and never touches the surface.  The transport leaves the electromagnetic cannon at 8.9 $km/s$.  It cruises out of the atmosphere and uses a small rocket engine to get into Earth's orbit.

A star tram system would cost about \$100 Billion.  It would launch about 150,000 tons of payload per year into orbit.  The transports experience 30 $g$ acceleration, thus they can not be used to transport passengers or fragile cargo.  Nevertheless, they can transport both sturdy cargo and fuel \cite{StarTram2010,StarTram2010A}.

\subsection{Electric propulsion in space}
An enormous cost reduction for deep -- space missions is possible if one could  lower the mass of fuel necessary for a propulsion within Deep Space.  Even though the fuel cost on Earth is small enough, in the LEO the cost of fuel and fuel tanks mainly consists of the cost for bringing these materials into the orbit, which is at least \$1,400 per $kg$.

The energy and fuel requirements for space missions are generally given in terms of
the change in velocity, $\triangle v$, of the spacecraft needed to accomplish a given mission \cite[p.9--11]{spaceprop}.  The amount of fuel needed for a maneuver with $\triangle v$ is given by the Tsiolkovsky rocket equation \cite[p.9]{spaceprop}
\be
\label{5.01}
\frac{m_0}{m_1}=\exp \left( \frac{\triangle v}{v_0} \right),
\ee
where $m_0$ is the mass of the rocket with the fuel, $m_1$ is the mass of the vehicle after the fuel has been consumed, and $v_0$ is the exhaust velocity of the rocket engine.  The highest $v_0$ for a storable liquid rocket propellant  (hydrazine and liquid fluorine) is $4.22\ km/s$ \cite{LRP}.  Thus, in order for $1\ kg$ of a payload and a fuel tank to perform a mission with
$\triangle v=10\ km/s$ one needs
\be
\label{5.02}
m_0=1\ kg \exp \left( \frac{10.0\ km/s}{4.22\ km/s} \right)-1=10.7 kg
\ee
of fuel.  Even if fuel tanks are very light and some of them are discarded along the way, we require at least $11\ kg$ placed in LEO to accomplish the mission for $1\ kg$ payload. Since placing $1\ kg$ into LEO costs at least \$1,400, it would cost at least $11 \times \$1,400 \thickapprox \$15,400$ per kilogram of payload at the mission end.

An obvious way to deliver greater mass over space routes, demanding large
$\triangle v$, is to use engines with exhaust velocities of tens of kilometers per second.  Several types of electrical thrusters with large exhaust velocities exist.  We briefly outline several types of these thrusters.  The \textbf{ion thruster} accelerates ions by an electric field.  It was proposed by Konstantin Tsiolkovsky in 1911 \cite{ionhist}.  The engine diagram from \cite[p.5]{ion1} is presented on Fig.2 below:
  \begin{center}
  \includegraphics[width=10cm]{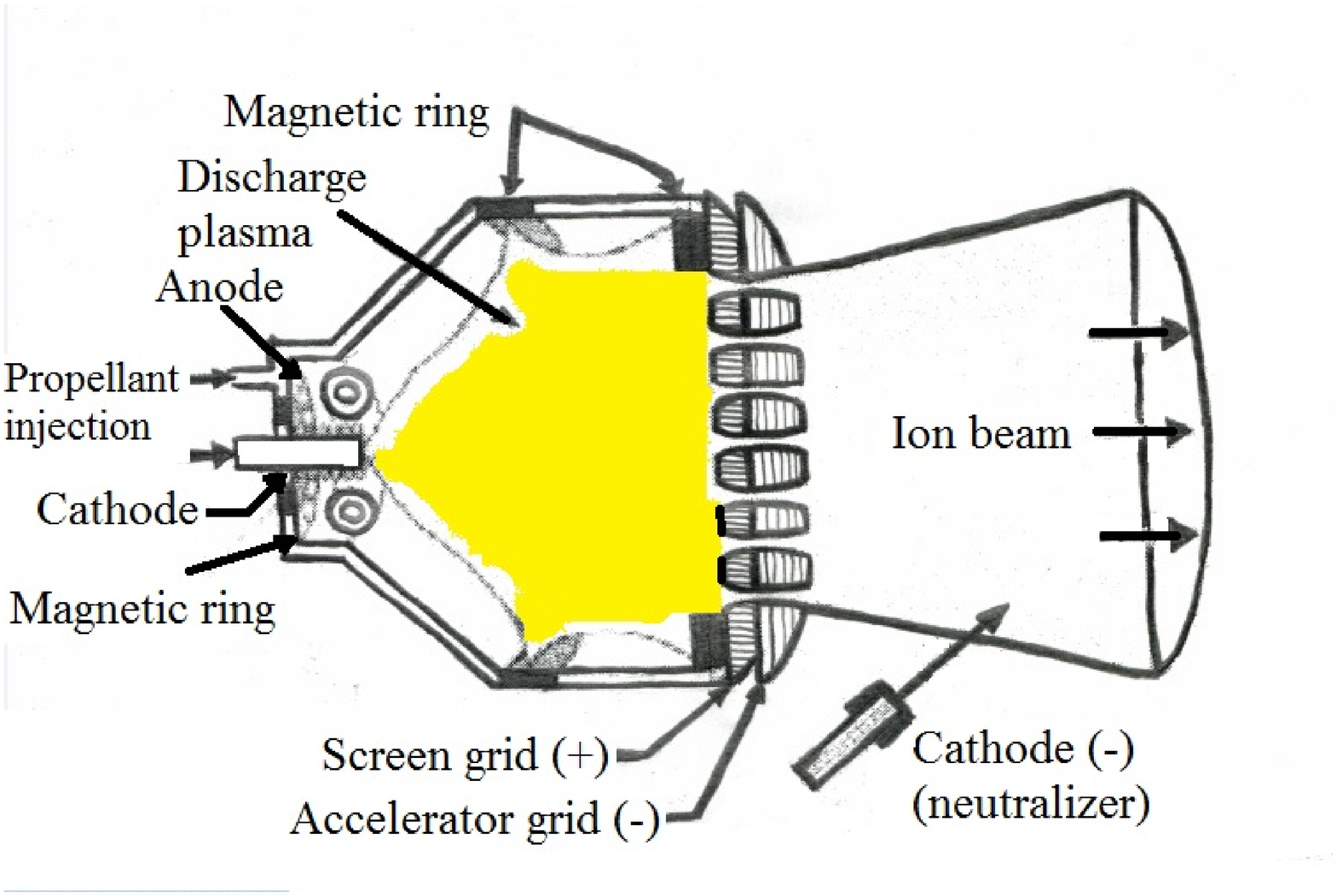}
  \captionof{figure}{Ion thruster}
  \end{center}
\textbf{Hall effect thruster} is described in \cite[p.6]{ion1}:
\begin{quote}
First, the propellant is ionized.  Then, the positively charged ions and negatively charged electrons are accelerated in the same direction by electric fields.  The electrons have the same velocity as the ions, and thus they have much less mechanical energy.

  Hall-effect thrusters typically provide thrust in the region of $100\ mN$ and a specific impulse of $1500-1700\ s$. Their principle is to use a rotating plasma of electrons to ionize a propellant injected though an anode. The configuration of the thrusters is such that a radial magnetic field is generated, via inner and outer magnetic coils. An axial electric field is also generated, and the combined effect of these fields generates the Hall effect, which confines the electrons to move in a direction perpendicular to $E$ and $B$, therefore setting up the azimuthal rotation. The ions are too heavy to be significantly affected by the magnetic fields. They accelerate axially under the influence of the electric field and exit the engine at high velocity, producing thrust.
\end{quote}

\hskip.5cm Another type of thruster is the \textbf{magnetoplasmadynamic thruster} \cite{MPT1}. Magnetoplasmadynamic thruster passes current through a propellant and uses Lorentz force to expel it at 15 to 60 $km/s$.  Such a thruster  generally has high power of $100$ -- $500$ $kW$.  Its construction is quite simple \cite[p.15]{MPT2} and schematically presented on Fig. 3.
  \begin{center}
  \includegraphics[width=5cm]{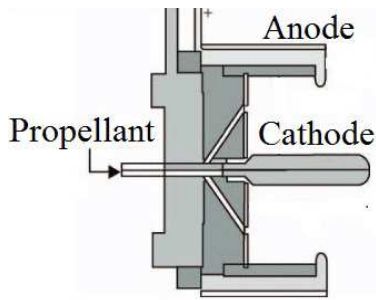}
  \vskip0cm
  \captionof{figure}{Magnetoplasmadynamic thruster}
  \end{center}
Since this type of thruster requires hundreds of kilowatts of electric power, it has not been used in space yet.  This is an example when the simplest idea is the hardest to implement.

Variable specific impulse magnetoplasma rocket (VASIMR) `` is an electromagnetic thruster for spacecraft propulsion. It uses radio waves to ionize and heat a propellant. Then a magnetic field accelerates the resulting plasma to generate thrust (plasma propulsion engine)" \cite{Vasya}.

An \textbf{arcjet} is a thruster in which the propellant is heated by an electric discharge.  The propellant is almost always Hydrogen.  The main problem with arcjets is that many of them have exhaust velocity below 10 $km/s$.

For a thruster with power $P$, exhaust velocity $v_0$, and efficiency $\eta$, the thrust force is given by the formula (see Subsection 3.1):
\be
\label{5.03}
F=\frac{2 P \eta}{v_0}.
\ee
Characteristics of several ion and Hall effect thrusters, which have been used in space, are given in \cite[p.55]{ion1}.  Those for magnetoplasmadynamic (MPD) thrusters are given in \cite[p.19]{MPT2}.  Those for VASIMR thrusters are given in \cite[p.7]{Vasya1}.

In Table 1 below, we describe the performance of several tested thrusters.  Column 1 contains the engine type.  Column 2 contains the engine name.  "Fakel'' means torch and "Energiya'' means energy in Russian.  Column 3 contains the engine propellant.  Xenon is unreasonably rare and expensive.  Column 4 contains the engine power -- we are mostly interested in engines with at least 300 kW of power. Column 5 contains the engine mass.  Column 6 contains the engine exhaust velocity.  All the listed exhaust velocities are within the acceptable range.  Column 7 contains the engine efficiency.
\begin{center}
\begin{tabular}{|l|l|c|c|c|c|c|}
  \hline
  Engine type & Engine  & Propellant  & Power    & Thruster  &$v_0$ $km/s$   & Efficiency \\
         & name    &             & $kW$     & mass (kg)     &            &            \\
  \hline
  Ion    & T6      & xenon       & 6.8     &            &46            & .68  \\
  Ion \cite{Next}     & NEXT    & xenon       & 6.9     &  13.5      &41            & .7  \\
  Ion    & HiPEP   & xenon       & 30      &            &87            & .8  \\
  \hline
  Hall   & SPT-140 & xenon       & 5       &            &49            & .55 \\
  \hline
  VASIMR \cite[p.7]{Vasya1}      & VF-150  & argon       & 150     &   1,000         & 29            & .61 \\
  VASIMR \cite{Vasya2}& VX-200   & argon & 200     & 620  & 30 -- 50  & .6 \\
  VASIMR \cite[p.7]{Vasya1}      & VF-400  & argon       & 400     &   1,000           & 49            & .73 \\
  \hline
  MPD \cite{MPT2,MPT3}   & Fakel   & lithium     & 300 -- 500 &  1000       & 34 -- 44 & .4 -- .6 \\
  MPD \cite{MPT2,MPT3}   & Energiya & lithium    & 500     &     1000       & 44            & .55 \\
  MPD \cite[p.34]{MPDynamic}   &          & sodium     & 25  &    & >18  & 0.4 \\
  MPD \cite[p.34]{MPDynamic}   &          & potassium  & 25  &    & >12  & 0.3 \\
  MPD \cite[p.20]{MPDT02}    &         & $H_2$   & 122     &                & 59            & .34 \\
  MPD \cite[p.101]{NucSpace01} &       & $H_2$   & 3,630   &                & 92            &
  .47 \\
  MPD \cite[p.449]{MPDT08} &       & argon       & 3,700   &                & 47            & .30 \\
  MPD \cite[p.450]{MPDT08} &       & $H_2$    & 3,700   &                & 210           & .60 \\
  \hline
  ArcJet \cite{ArcJet2}    & RAD X-1 & $H_2$ & 216 &     & 21.7   & .35 \\
  ArcJet \cite{ArcJet2}    & R-2     & $H_2$ &  30 &     & 14.8   & .36 \\
  ArcJet \cite{ArcJet4}    & HIPARC  & $H_2$ & 410 &     & 20     & .29 \\
  \hline
\end{tabular}
\captionof{table}{Electric thrusters \label{T.02}}
\end{center}
Argon and lithium are readily available, while xenon is one of the rarest elements on the Earth.  The efficiency of VX-200 engine for different propellants is presented in \cite[p.8]{Vasya3}:
  \begin{center}
  \includegraphics[width=15cm]{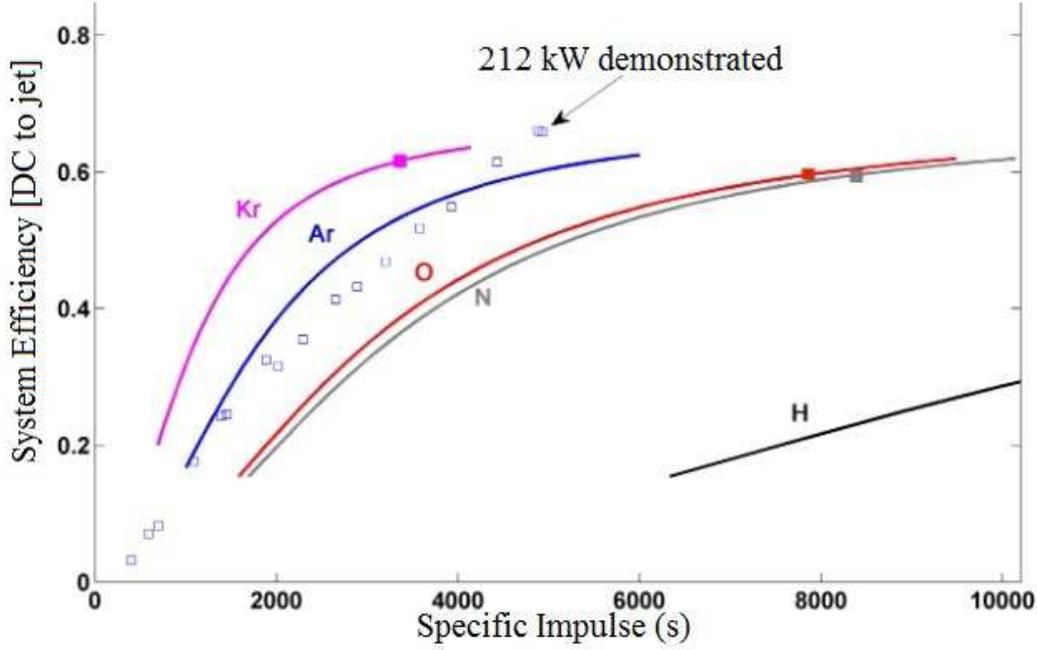}
  \vskip0cm
  \captionof{figure}{System efficiency of the VX-200 engine as a function of the specific impulse.}
  \end{center}
As seen from Fig.4, krypton gives considerably better performance than argon, especially for lower specific impulses.  Nevertheless, argon gives efficiency over 57\% for specific impulses of over $4,000\ s$ corresponding to exhaust velocity of $40\ km/s$.  Thus, argon should be used, since krypton is very rare and expensive.

Using the aforementioned thrusters, missions with considerable $\triangle v$ are possible.  Using formula (\ref{5.01}), we calculate the amount of fuel needed to deliver $1\ kg$ of payload into space to perform a $\triangle v=10\ km/s$ mission
\be
\label{5.04}
m_0=1\ kg \exp \left( \frac{10.0\ km/s}{44\ km/s} \right)-1\ kg=0.26\ kg.
\ee
Even though the engine and power generation system take considerable weight, with such engines deep space missions are possible.

Sodium should be further investigated as MPD propellant.  Sodium MPD thrusters may be important for colonization of Mercury, Mars, and the Asteroid Belt.  First, sodium is widely available throughout the Solar System.  Second, sodium is much more storable than hydrogen.

\subsection{Thermal propulsion}
\hskip.5cm   A thermal rocket contains a heat source, through which the propellant is passing.  The heat source can consist either of a nuclear reactor or of a metal container within a focus of a parabolic mirror.  The heated propellant is expanded and expelled via an exhaust nozzle.

Within the Asteroid Belt and Outer Solar System, water and ammonia are plentiful.  These substances can be used as thermal rocket propellant there.  In the vicinity of Mercury, only hydrogen can be used as a high specific impulse propellant.
Solar thermal rockets can heat hydrogen to 1,900 $^o$C \cite[p.205]{Advanced2020}.
Nuclear reactor can heat hydrogen to 2,470 $^o$C to 2,670 $^o$C  \cite[p.9]{NTh1}.

We use Rocket Propulsion Analysis (RPA) program \cite{RPA} to calculate performance of thermal rockets using hydrogen propellant.  The results are collected in Table \ref{T.03} below.
The first  column is the rocket chamber temperature.
The second column is the rocket chamber pressure.
The third  column is the nozzle expansion ratio.
The fourth  column is the exhaust velocity.

\begin{center}
\begin{tabular}{|r|r|r|r|r|r|}
  \hline
  $T,\ ^oK$   & $P,\ atm$& EER        & $v,\ m/s$\\
  \hline
  2,100       & 20       & 120        & 7,330    \\
  2,200       & 20       & 120        & 7,520    \\
  2,300       & 20       & 120        & 7,710    \\
  2,500       & 20       & 120        & 8,110    \\
  2,700       & 20       & 120        & 8,520    \\
  2,900       & 20       & 120        & 9,010    \\
  3,000       & 20       & 120        & 9,270    \\
              &          &            &          \\
  \hline
\end{tabular}
\begin{tabular}{|r|r|r|r|r|r|}
  \hline
  $T,\ ^oK$   & $P,\ atm$& EER        & $v,\ m/s$\\
  \hline
  2,500       & 2        & 120        &  8,250    \\
  2,700       & 2        & 120        &  8,820    \\
  2,900       & 2        & 120        &  9,540    \\
  3,000       & 2        & 120        &  9,950    \\
  \hline
  2,500       & 0.5      & 120        &  8,440    \\
  2,700       & 0.5      & 120        &  9,210    \\
  2,900       & 0.5      & 120        & 10,220    \\
  3,000       & 0.5      & 120        & 10,810    \\
  \hline
\end{tabular}
\captionof{table}{Performance of different propellants in thermal rocket \label{T.03}}
\end{center}

Liquid hydrogen thermal rockets give have much higher exhaust velocity than any chemical rockets.  A problematic issue related to using liquid hydrogen is that it must be stored below 20 $^o$K.  Hydrogen containers are discussed in Subsection 3.5 below.

Thermal propulsion uses much less energy than ionic propulsion.  First, it does not need to convert thermal energy into electrical energy.  Second, the lower exhaust velocity reduces cost per unit thrust, even though it greatly increases fuel consumption.

\section{The stages of evolution of GSRF}
In this section we briefly describe the stages of evolution of a guided self replicating factory.

\subsection{Stage 1: exploration}

The stage will begin with very detailed exploration of available resources in the Solar System.  The Moon will be explored by robotic probes remotely controlled from Earth.  Exploration of Mercury, Mars, and the Asteroid Belt will be more difficult as it will require missions with astronauts as well as robots.  Robotic probes guided by astronauts will make detailed maps and assessments of mineral resources available on Mercury, Mars, and the Asteroid Belt.  These probes will also access possibilities for construction.

As the data is gathered in situ, back on Earth, experimental and theoretical scientists as well as graduate students will be designing the further stages of development of colonies in minute detail.  It is likely that different companies will have competing designs.  Design of any space system takes thousands to tens of thousands of expert-years \cite[p.116]{SLVD}.

\subsection{Stage 2: setup of Lunar colony}

In the author's opinion, the colonization of the Solar System will start with the Moon.
The Moon is close to Earth, and it has resources which are extremely important for space transportation.
Detailed feasibility studies for Lunar colonization have been published since 1970s \cite{LunCol,LunCol1,LunCol2,LunCol3}.

Below, we briefly describe some Lunar resources and discuss their role in colonization of the rest of the inner Solar System.
The first Lunar resource is hydrogen in the form of water ice.
Lunar poles contain at least 430 million tons of ice \cite[p.326]{LunarIce}.
Other sources estimate the ice deposits at lunar poles as 660 million tons \cite[p.368]{LunaResources1}.  In most polar craters which contain ice, water makes up 2.5\% of the top 1.5 meters of regolith.  Based on the data above, the Moon contains 48 million tons to 73 million tons of easily accessible hydrogen at the Lunar poles.
The second Lunar resource is abundance of metal ores.  Chemical composition of different parts of lunar surface has been summarized in \cite{LunarAu}.
Lunar surface contains
24\% to 31\% aluminum oxide,
14\% to 18\% calcium oxide,
6\% to 10\% iron oxide,
3\% to 8.6\% magnesium oxide,
0.3\% to 0.6\% sodium oxide,
0.2\% to 0.6\% titanium oxide, and
0.09\% to 0.18\% chromium oxide \cite[p. 400]{LunarAu}.
The third Lunar resource is the abundance of minerals useful for glassmaking.

Having listed lunar resources, we describe the setup of a lunar GSRF below.
First, an electric power station would be brought to the Moon from Earth.
Solar panels brought from Earth and covered with glass manufactured on the Moon would make a good power station.
In 2020, Renogy, Inc. sells flexible solar panels.  These panels cost under \$2 per watt or \$300 per $m^2$.  These panels have produce 175 $W/m^2$ under 1,000 $W/m^2$ irradiance.  They have specific mass of 2.8 $kg/m^2$ \cite{Renogy}.  Some lightweight cells have specific power of 948 $W/kg$ to 1,181 $W/kg$ \cite{LwCell1}.  These cells are much more expensive than regular photovoltaic cells.
The highest specific power for any solar cell commercially available in 2017 is 700 $W/kg$ for III-V triple junction module.  This module has 25\% efficiency and 350 $g/m^2$ areal density.  These modules cost \$25,000 per $m^2$, which \$100 per Watt and \$71,000 per $kg$ -- more expensive than gold \cite{LwCell2}.  Hopefully, by the time colonization of the Solar System begins, these prices will decrease.
Solar panels can be manufactured on the Moon from solar cells delivered from Earth and cover manufactured from lunar material.  Even in 1989, solar cells not assembled into panels and without cover had specific mass of 1,700 $W/kg$ \cite{LunarPV3}.  Some researchers claim that solar cells can be manufactured on the Moon.  Nevertheless, production of solar cells involves deposition of submicron thick semiconductor layers.  This technology may be too complicated for a lunar base \cite{LunarPV4,LunarPV1}.
A transparent glass covering for solar cells can be produced from lunar regolith after iron, nickel and chromium ores have been removed from it \cite[p.13]{MoonGlass1}.

Second, a material production system would be brought to the Moon from Earth.  Magnesium, aluminum, steel, titanium, and sodium would be manufactured from the resources listed above with the use of electric energy.  Vast amounts of hydrogen would also be produced.  Most of this hydrogen will be liquefied and stored in depots in space and on the Moon.

Third, a metal shaping and assembly system will be delivered to the Moon.  This system will be carefully monitored and controlled by tens of thousands of computers on Earth.

Fourth, a transport link would be set up between Moon and Earth.  Rockets using liquid hydrogen and oxygen will transport payloads between the lunar base and orbit around the Moon.  Chemical hydrogen-oxygen rockets as well as thermal rockets using only liquid hydrogen will transport payloads between orbit around the Moon and Low Earth Orbit(LEO).
Even though hydrogen is plentiful on Earth, in space it is an important and rare resource due to delivery cost.  Depots containing hydrogen manufactured on the Moon would be set up in depots in orbit around the Moon, LEO, and deep space.  Propellant depots in LEO would serve as refueling stations for spacecraft going to the Moon.  Propellant depots at Lunar Orbit would serve as refueling stations for spacecraft landing on the Moon as well as those returning to Earth.

LEO and deep space propellant depots can also serve as refueling stations for spacecraft travelling from Earth to other destinations in the inner Solar System.  That would greatly reduce the payload which must be delivered to LEO in order to deliver 1 $kg$ of payload to these destinations.  Colonization of the Solar System would be impossible without availability of propellant manufactured on the Moon.

\subsection{Stage 2.5: setup of one or more colony in inner Solar System}

During this stage, GSRF is built on Earth, transported to Mercury, Mars, or Asteroid Belt and assembled there.  Astronauts would have to accompany GSFR. Transportation of equipment and astronauts would be assisted by the use of hydrogen propellant manufactured at the lunar base.  In the author's opinion, this stage will be the most expensive and risky part of inner Solar System colonization.

% electric power station
First, an electric power station would be set up.
On Mercury, it would be a solar thermal electric power station.
On Mars, it would be a nuclear power station.
On the Asteroid Belt, the power station can be either solar thermal or nuclear.
All parts of electric power station will be brought from Earth.

% material production system
Second, a material production system would be set up.
Metal production system will manufacture metals for machines and structures, liquid hydrogen and oxygen for propulsion, and rubber for insulation.
On Mercury and Mars, metals would be obtained from the ores.  These ores will be gathered by mining robots.
On the Asteroid Belt, native metals are widely available.
Once again, all parts of this plant will be delivered from Earth.

An important part of material production system on Asteroid Belt would separate platinum group metals from metal asteroid material.  These metals would be delivered to Earth.  Production of platinum group metals would finance expansion of space colonization.

% material shaping and assembly system
Third, material shaping and assembly system will be set up.
This system will manufacture simple metal, plastic, and rubber details.
Below, we briefly describe some manufactured items.
Steel building blocks would be used by the astronauts to build more extensive and comfortable habitats for themselves.
Bolts, nuts, screws, metal bars, frames, wires, and details of other shapes would be used to expand the factory.
Polished aluminum mirrors would be used in concentrated solar power.
Tubes would be used in to expand the electric power station's heat sources and heat sinks.

% space transportation system
Fourth, space transportation system will be set up.
Initially, spaceships will be built on Earth.
Fuel for refueling spaceships will be produced in space colonies from the beginning.  Initially, most hydrogen will be produced on lunar colonies.
Hydrogen can be used as propellant for thermal nuclear rockets and magnetoplasmadynamic thrusters.  Later, hydrogen will be produced in large quantities on Mercury, Mars, and Asteroid Belt.
Sodium will also be produced in large quantities.  It will be used for magnetoplasmodynamic thrusters.

% exponential growth
Replication and expansion of the colony from in situ resources will take place during setup stage.  Nevertheless, the rate of replication will be much slower than at the next stage.  In the author's opinion, the Setup Stage should take about 10 years.

\subsection{Stage 3: exponential growth}

GSRF will enter Exponential Growth Stage with full set of components needed for complete self-sustained growth.
%% electric power station
Electric power plants will generate vast amount of electric energy.  On Mercury, the main energy source will be concentrated solar power.  On Mars and Asteroid Belt, it will be nuclear power.  This energy will power all other plants.

% material production system
On Mercury and Mars, mining machines will deliver metal ores, graphite (on Mercury), water ice, and dry ice (on Mars) to the factory.
Ore benefaction machines will concentrate metals and other useful compounds.
Metal extraction plants will use electric energy and minerals to produce metals.
Chemical plants will use electricity to convert ice and graphite or dry ice into propellant, rubber, plastic, and other chemicals.
On Asteroid Belt, both steel and hydrocarbons are available in native state.
Chemical plants will modify and purify metals and hydrocarbons.  Chemical plants on Asteroid Belt will also extract platinum group metals.  These metals will be sold on Earth in order to finance space colonization.

% material shaping and assembly system
Metal working plants will shape metals and plastics into building blocks and machine parts.
Astronauts assisted by machine building robots will produce machines for all aforementioned plants.
The process described above will neither be fully automated nor completely independent of complex machines, robots, and computers manufactured on Earth in great quantities.  Thousands of astronauts will be arriving at Mercury, and some will be returning to Earth.  Total payload from Earth may be greater than during the previous stage.

% space transportation system
Transportation from Earth will also be bootstrapped by resources manufactured on one or more inner Solar System colonies.  Deep spaceships will be refueled by propellant manufactured at the colony.  As the colony expands, parts of deep spaceships or whole deep spaceships will be manufactured there.  This will reduce the cost of delivering astronauts and machinery from Earth to Mercury, Mars, and Asteroid Belt.

% exponential growth
After several doubling periods, both qualitative and quantitative changes will occur in the colony.  The colony will become less materially dependent on Earth.  Specialized factories will be set up producing robotic machinery of increasing complexity.  The number of astronauts brought to colonies will greatly increase.  There will be a great need for astronauts to operate machinery.  Even though the mass and power of machinery controlled by each astronaut will grow, this growth will be less rapid than the growth of the colony itself.
Energy and material resources will be used to build comfortable accommodations for astronauts.  Hydroponic food production will enable astronauts to eat a healthy diet \cite{HPonic}.

\subsection{Stage 4: full civilization}
In this subsection, we use the present tense in order to demonstrate the great potential of Solar System colonization.
After the colony undergoes about 10-12 doubling periods, it changes both quantitatively and qualitatively.  It contains tens of thousands of astronauts and several million tons of structure and machinery.  Electric power stations have cumulative power of 20 $GW$ to 50 $GW$.
The original colony has been built primarily in one place -- Mercury, Mars, Ceres, or other asteroids.  Now, offshoots of the original colony are throughout the inner Solar System.
Originally, the colony consumed more resources, than it produced.  Now the colony produces much more resources than it consumes.  It produces platinum-group metals, silver, and gold which are sold on Earth.
Originally, the colony brought most machinery from Earth.  Now, almost all machinery is manufactured in situ.

Originally, no astronaut viewed Mercury, Mars, or the Asteroid Belt as their home.  Now, many do.  Originally, astronauts had only makeshift habitats.  Now they have comfortable living quarters in moderate-sized settlements with hydroponic farms growing healthy food.  First people are born in space.  Politically, different parts of the colony are owned by different companies and Nations.  Most people on Earth have stocks from asteroid mining companies. Space Law and Space Commerce are emerging disciplines.

Expansion of Space Colonies does not stop at this stage.  On the contrary, almost unlimited resources of solar energy, metals, minerals, carbon, hydrogen, and oxygen allow unlimited expansion.  An exponentially growing number of people are calling different parts of the Solar System their home.

\subsection{Stage 5: toward Dyson Sphere}

After the colony undergoes 15-18 more doubling periods from the beginning of Stage 4, it contains about a trillion tons of structure and machinery.  The gross output of all electric power stations is 5 $PW$ to 10 $PW$, where $PW$ denotes a Petawatt or $10^{15}\ W$.  The population of Space Civilization is now a billion people.  New generation is born in Solar System colonies, and the younger generation is rapidly emigrating from Earth.  Space Civilization is now fully self-sufficient.

Social and political development of the True Space Age civilization is beyond our imagination.  Civilization described in this stage frequently appears in Science Fiction \cite{SFict1,SFict2}.  In reality, it is impossible to foresee problems and opportunities which will exist in this civilization.

Solar System civilization may continue growing until all resources are exhausted.  After the colony undergoes 30-33 more doubling periods from the beginning of Stage 5, it becomes a Dyson Sphere -- a multitude of space habitats orbiting Sun \cite{Dyson}.  Mass, power, and population of Dyson Sphere have been estimated by various researchers since the 1960s.  These estimates are based on resources available in the Solar System.  Population of about $10^{16}$, which is over a million times modern World population would be supported at very high living standards \cite{skymine}.  Dyson Sphere would contain at least $10^{19}\ tons$ of structure and machinery.  This material would come from Asteroid Belt, Kuiper Belt, Mercury, and satellites of gas giants \cite{crc, ABM, MKuiper}.  Total power output of electric stations would be at least $3 \cdot 10^{24}\ W$ -- which is just 0.75\% of Sun's total radiative power.  Even though these numbers seem extraordinary, most scientists who have studied the issue since the 1960s agree on them.

\section{Conclusion}
Colonization of the Solar System can be accomplished by the use of a \textbf{guided self-replicating factory}.  This factory would consist of four elements.  Electric power station would power the whole factory.  Material production system would gather ores or native metals and produce steel, aluminum, plastic, and propellant such as liquid hydrogen.  Material shaping and assembly system would shape metal and plastic into details and assemble them into machines.  These machines would constitute an expansion of the colony.  Space transportation system would use fuel manufactured in situ in order to facilitate transportation of machinery and astronauts from Lower Earth Orbit (LEO) to the colony.

Any colony located on Mercury, Mars, or Asteroid Belt would be guided and intensely monitored by astronauts.  A lunar colony would be remotely controlled from the Earth.  Building a fully robotic and independent, self replicating factory is beyond current technological capabilities.  It is possible that such a factory would be built only in the very distant future by a Dyson Sphere civilization, which would use the vast resources of the whole Solar System.

The GSFR will not exactly replicate itself.  Nevertheless, it will double during a time period called a doubling time.  During doubling time, the mass of machinery and structure doubles due to material collected, processed, and assembled in situ.  At the same time, electric power doubles.  The number of astronauts needed to work on GSFR increases by a factor a little less than 2.

After 10-12 doubling times, the space colony becomes a settlement with tens of thousands of astronauts, millions of tons of structure, and machinery, and electric power output of 20 $GW$ to 50 $GW$.  By that time, there should be more than one colony within the Solar System.  After another 15-18 doubling times, True Space Age will begin.  A significant fraction of world population will be living in space habitats.  The civilization within the Solar System will continue to grow until it becomes a Dyson Sphere civilization.  Dyson Sphere would support a population of $10^{16}$ at very high standards of living.  This population would be supported by at least $10^{19}$ tons of structure and machinery.  Global production of electric power would be at least $3 \cdot 10^{24}\ W$.

Setting up original colonies on the Moon and in the inner Solar System would be a difficult and expensive task.  During the first stage of growth, this colony may need more expensive equipment from Earth.  It is likely that the overall expense for Solar System colonization would be 500 Billion dollars or more.  Nevertheless, this would be an enormously important step in transforming Humankind into a Dyson Sphere civilization.


\begin{thebibliography}{2000}
\bibitem{skymine} Lewis, J.S. \emph{Mining the Sky: Untold Riches from the Asteroids, Comets, and Planets,} Reading, Mass: Addison-Wesley Pub. Co, 1996.
\bibitem{kard} Kardashev, N. S., Transmission of information by extraterrestrial civilizations, \emph{Soviet Astronomy}, \textbf{8}(2), Sept.-Oct. 1964.
\bibitem{E1} Smil, V., \emph{Energy and Civilization: A History}, The MIT Press, Cambridge, MA, 2017.
\bibitem{crc} Lide, D. R., Editor,  \emph{CRC Handbook of Chemistry and Physics,} 84$^{\text{th}}$ Edition, CRC Press. Boca Raton, Florida, 2003.
\bibitem{wes} OCDE, OECD,  \emph{World Energy Statistics 2016}, OECD Publishing, 2016. Internet resource.
\bibitem{Dyson} Beech, M., \emph{Terraforming: The Creating of Habitable Worlds,} New York: Springer, 2009.
\bibitem{ABM} Pitjeva, E.V., High-precision ephemerides of planets—EPM and determination of some astronomical constants, \emph{Solar Syst. Res.} \textbf{39}(3), p. 176–186, (2005).
\bibitem{Asteroids} Binzel, R.P., Gehrels, T., Matthews, M.S., \emph{Asteroids II.} Tucson: University of Arizona Press, 1989.
\bibitem{Psycho} Al Conrad, P. I., Adamkovics, M.,  Kleer K., Males, J.R., Morzinski, K.M.,  Close, L., Kaasalainen, M., Viikinkoski, M., Timerson, B.,  Reddy, V., Magri, C., Nolan, M.C., Howell, E.S.,  Benner, L.,  Giorgini, J.D., Warner, B.D  and Harris, A.W., Radar Observations and Shape Model of Asteroid 16 Psyche, \emph{Icarus,} \textbf{281}, p. 388 -- 403, (2017).
\bibitem{Kuiper}  Blondel, P., Mason, J., \emph{Solar System Update,} Springer-Verlag, Berlin, 2006.
\bibitem{MKuiper}  Pitjeva, E.V.,  Pitjev, N.P., Mass of the Kuiper belt, \emph{Celestial Mechanics and Dynamical Astronomy,} \textbf{130}(9), 2018.
\bibitem{CruiseShips}  Smith, P.C., \emph{Cruise Ships the Small Scale Fleet: A Visiual Showcase,} Pen and Sword, Havertown, 2014.
\bibitem{Tsialkovski} Tsiolkovski, K., Tikhonravov, M.K., \emph{Works on Rocket Technology,} Washington, D.C.: National Aeronautics and Space Administration, 1965.
\bibitem{habitats} O'Neill, G.K. and Reynolds, G., Habitats in Space, \emph{The Science Teacher,} \textbf{44}(6), p. 22-26, 1977.
\bibitem{1974} O'Neill, G.K., The Colonization of Space, \emph{Physics Today}, \textbf{27}(9) p. 32-40, 1974.
\bibitem{cost}  Hickman, J.,  The Political Economy of Very Large Space Projects, \emph{Journal of Evolution and Technology.} Jetpress.org. 4, November, 1999.
\bibitem{planz01} Klotz, I, Tech billionaires bankroll gold rush to mine asteroids, \emph{Reuters,} April 24 , 2012.\\ <http://www.reuters.com/article/2012/04/24/us-space-asteroid-mining-idUSBRE83N06U20120424>
\bibitem{SRM02}  Chirikjian, G.S.,\emph{ An Architecture for Self-Replicating Lunar Factories,} Department of Mechanical EngineeringJohns Hopkins University, 2004.
\bibitem{SRM1}  Sipper, M., Fifty Years of Research on Self-Replication: An Overview, \emph{Artificial Life,} \textbf{4}, pp. 237–257, 1998.
\bibitem{SRM2}  Ellery, A., Building Physical Self-Replicating Machines, \emph{The European Conference on Artificial Life,} Lyon, France, 2017.
\bibitem{SRM3}  Sayama, H., Von Neumann's Machine, \emph{in the Shell: Enhancing the Robustness of Self-Replication Processes, in Artificial Life VIII,} MIT Press, pp. 49-52, 2002.
\bibitem{SRM4}  Chirikjian, G.S., Moses, M.S., Yamaguchi, H., Towards cyclic fabrication systems for modular robotics and rapid manufacturing, \emph{Proceedings of the 2009 IEEE/RSJ international conference on Intelligent robots and systems, October 2009,}  pp. 1478–1483, 2009.
\bibitem{SRM8}  Boston, P. J., Todd, P., McMillen, K. R., Robotic Lunar Ecopoiesis Test Bed: Bringing the Experimental Method to Terraforming, \emph{Aip Conference Proceedings,} pp.975-983, 2004.
\bibitem{SRM9}  Chirikjian, G.S., \emph{An Architecture for Self-Replicating Lunar Factories,} \emph{NIAC Phase I Award: October 1, 2003 - March 31, 2004,} 2004.
\bibitem{SRM03}  Toth-Fejel, T.,\emph{ Modeling Kinematic Cellular Automata Final Report,} NASA Institute for Advanced Concepts Phase I: CP-02-02, General Dynamics Advanced Information Systems Contract \# P03-0984, 2004.
\bibitem{SciFi1}  Cyrano, B.S., Strachan, G., \emph{Other Worlds: The Comical History of the States and Empires of the Moon and the Sun,} London: New English Library, 1976.
%\bibitem{SciFi2}  Shelley, M., \emph{Frankenstein,} Tustin: Xist Publishing, 2015.
\bibitem{SciFi3}  Wells, H.G., \emph{War of the Worlds,} London: Legend Press, 2020.
\bibitem{Solar1831}  Etzler, J.A.,  \emph{The Paradise Within the Reach of All Men, Without Labour, by Powers of Nature and Machinery: An Address to All Intelligent Men,} London: J. Brooks, 1836.
\bibitem{Mechanical1841}  Etzler, J.A., \emph{The New World: Or, Mechanical System, to Perform the Labours of Man and Beast by Inanimate Powers, That Cost Nothing, for Producing and Preparing the Substances of Life,} Philadelphia: C.F. Strollmeyer, 1973.
\bibitem{GalCol1} Savage, M.T., \emph{The Millennial Project: Colonizing the Galaxy in Eight Easy Steps,} Boston: Little, Brown, 1994.
\bibitem{GalCol2} Finney, B.R., Jones, E.M., \emph{Interstellar Migration and the Human Experience,} Berkeley: University of California Press, 1986.
\bibitem{Boot1}  Metzger, Philip T., Muscatello, A., Mueller, R. P. and Mantovani, J., Affordable, rapid bootstrapping of space industry and solar system civilization, Journal of Aerospace Engineering, \textbf{26}, pp.18-29, 2013.
\bibitem{Boot2}  Mackenzie, B.A. Bootstrapping Space Resource Utilization with Tethers, Regolith Rockets and Micro Rovers, \emph{Proceedings of the 5th International Conference on Space’ 96, Albuquerque, NM,} pp. 321–327, 1996.
\bibitem{BrAge}  Weeks, L.R., \emph{Early Metallurgy of the Persian Gulf: Technology, Trade, and the Bronze Age World,} Boston: Brill, 2004.
\bibitem{Ugol}   Alkire, J., \emph{Coal Energy: Putting Rocks to Work,} Minneapolis, Minnesota : Super Sandcastle, an imprint of Abdo Publishing, 2019.
\bibitem{ThIon1} Go, D.B., Haase, J.R., George, J., Mannhart, J., Nemanich, R., Nojeh, A., Wanke, R., Thermionic energy Conversion in the Twenty-first Century: Advances and Opportunities for Space and Terrestrial Applications, \emph{Frontiers in Mechanical Engineering,} \textbf{3}(13), 2017.
\bibitem{ThIon4} Alkasim, A., Usman, A.,   Maximum Conersion Efficiency of Thermionic Heat to Electricity Converters Using Pure Tungsten as the Emitter: a Theoretical Review, \emph{Global Journal of Pure and Applied Sciences,} \textbf{17}(1), pp.71-79, 2011.
\bibitem{ThIon3} Khalid, K.A.A., Leong, T.J., Mohamed, K., Review on Thermionic Energy Converters, \emph{IEEE Transactions on Electron Devices,} \textbf{63}(6), 2016.
\bibitem{PETE2}  Kribus, A., Rosenwaks, Y., Segev, G., Limit of efficiency for photon-enhanced thermionic emission vs. photovoltaic and thermal conversion, \emph{Solar Energy Materials \& Solar Cells,} \textbf{140}, pp.464–476, 2015.
\bibitem{PETE3}  Kribus, A., Segev, G.,  Solar energy conversion with photon-enhanced thermionic emission, \emph{Journal of Optics,} \textbf{18}(7), 2016.
\bibitem{PETE1}  Bargatin, I., Hardin, B.E., Howe, R.T., Melosh, N.A., Pianetta, P., Riley, D.C., Rosenthal, S.J., Schmitt,F., Schwede, J.W., Shen, Z.X., Sun, Y., Photon-enhanced thermionic emission for solar concentrator systems, \emph{Nature Materials,} \textbf{9}, pp. 762-767, 2010.
\bibitem{PETE7}  Kribus, A., Segev, G., Solar energy conversion with photonenhanced thermionic emission, \emph{Journal of Optics,} \textbf{18}, 2016.
\bibitem{PETE5} Alabastri, A., Cunha, J., Raja, W., Summerer, L., Versloot, T.W., Zaccaria, R.P., Zilio, P., \emph{Photon-Enhanced Thermionic Emission,} 2015.
\bibitem{GEff01}  \emph{Steam Turbines,} The Industrial Press, New York City, 1911.
\bibitem{Eg1}  Farokhi, S., \emph{Future Propulsion Systems and Energy Sources in Sustainable Aviation,} Newark: John Wiley \& Sons, Incorporated, 2020.
\bibitem{Eg2}  Stuart, S., \emph{Electrical (Generator and Electrical Plant): Modern Power Station Practice,} Elsevier Science, Amsterdam, Netherlands, 2013.
\bibitem{CS1}  Tyagi, H., Chakraborty, P.R., Powar, S., Agarwal, A.K., \emph{Solar Energy: Systems, Challenges, and Opportunities,} Singapore: Springer Singapore, 2020.
\bibitem{CS2}  Taylor, N., \emph{Solar Thermal Electricity: Technology Development Report,} Luxembourg : Publications Office of the European Union, 2019.
\bibitem{CS3}  Narducci, D., Bermel, P., Lorenzi, B., Wang, N., Yazawa, K., \emph{Hybrid and Fully Thermoelectric Solar Harvesting,} Cham, Switzerland: Springer, 2018.
\bibitem{SPwr06}  Yoder, G.L., \emph{Technology Development Program for an Advanced Potassium Rankine Power Conversion System Compatible with Several Space Reactor Designs,} Washington, D.C: United States. Department of Energy, 2005.
\bibitem{SPwr05a}  Mason, L.S., \emph{A Comparison of Brayton and Stirling Space Nuclear Power for Power Levels from 1 Kilowatt to 10 Megawatts,} National Aeronautics and Space Administration, NASA/TM—2001-210593, 2001.
\bibitem{CDioxTurb2}  Dennis, R., \emph{Overview of Supercritical Carbon Dioxide Based Power Cycles for Stationary Power Generation,} Presented to: ARPA-E Workshop on High Efficiency High Temperature Modular Power Utilizing Innovative Designs, Materials, and Manufacturing Techniques, 2017.
\bibitem{CDioxTurb3}  Almitani, K.H., Siddiqui, M.E., Energy Analysis of the S-CO$_2$ Brayton Cycle with Improved Heat Regeneration, \emph{Processes,} \textbf{7}(3), 2019.
\bibitem{CDioxTurb5}  Zhu, Q., \emph{Power generation from coal using supercritical CO$_2$ cycle,} IEA Clean Coal Centre, 2017.
\bibitem{Bray} Dyreby, J.J.,  \emph{Modeling the Supercritical Carbon Dioxide Brayton Cycle with Recompression,} 	University of Wisconsin--Madison, Madison, WS, 2014.
\bibitem{SPwr03}  George, J.A., \emph{Multi-Reactor Power System Configurations for Multimegawatt Nuclear Electric Propulsion,} NASA Technical Memorandum 105212, 1991.
\bibitem{SPwr05}  McGinnis, S.J., \emph{Space Nuclear Power Systems for Manned Mission to Mars,} Master’s Thesis, Naval Postgraduate School, 2004.
\bibitem{KTurbo1} Moor, B. L., Schnetzer, E., \emph{Three-stage Potassium Vapor Turbine Test,} Defense Technical Information Center, Ft. Belvoir, 1971.
\bibitem{KTurbo2} Fraas, A.P., Burton D.W., LaVerne M.E., Wilson, L.V., \emph{Design Comparison of Cesium and Potassium Vapor Turbine-Generator Units for Space Power Plants,} Oak Ridge National Laboratory, 1969.
\bibitem{KTurbo3}  Supak, K.R., \emph{Reduced Gravity Rankine Cycle System Design and Optimization Study with Passive Vortex Phase Separation,} College Station, Texas: Texas A \& M University, 2008.\\
    <https://pdfs.semanticscholar.org/1124/0c59b699caed7ef7b8df143c255fbaa5d310.pdf>\\
    Accessed Jan 20, 2020.
\bibitem{Stirling3} Berchowitz, D.M., Kim, S.Y., Specific Power Estimations for Free-Piston Stirling Engines, \emph{4th International Energy Conversion Engineering Conference and Exhibit (IECEC),} 26 - 29 June 2006, San Diego, California, 2006.
\bibitem{SPwr01}  George, J.A., Scott, J.H., Tarditi, A.G., \emph{Direct Energy Conversion for Low Specific Mass In-Space Power and Propulsion,} Proceedings of Nuclear and Emerging Technologies for Space 2013.
\bibitem{Centrifuga}  Ghose, A.K., Bose, L.K., \emph{Mining in the 21st Century: Quo Vadis?,} New Delhi: Balkema, 2004.
\bibitem{MercChComp}  Lawrence, D. J., Peplowski, P. N., Beck, A. W., Feldman, W. C., Frank, E. A., McCoy, T. J., Nittler, L. R., Solomon, S. C., Compositional Terranes on Mercury: Information from Fast Neutrons, \emph{Icarus,} 281 pp. 32-45, 2017.
\bibitem{MercC4}  Weider, S.Z., \emph{Petrology and Geochemistry of Mercury,} Oxford University Press, Oxford, UK, 2018.
\bibitem{MercC1}  Peplowski, P. N., Klima, R. L., Lawrence, D. J., Ernst, C. M., Denevi, B. W., Frank, E. A., Goldsten, J. O., Murchie, S.L., Nittler, L.R., Solomon, S. C., Remote sensing evidence for an ancient carbon-bearing crust on Mercury, \emph{Nature Geoscience,} \textbf{9}(4), pp. 273-276, 2016.
\bibitem{MercC3} Vander, K.K.E., McCubbin, F.M.,  Exotic Crust Formation on Mercury: Consequences of a Shallow, Feo-Poor Mantle, \emph{Journal of Geophysical Research: Planets,} \textbf{120}(2) pp.195-209, 2015.
\bibitem{MH2O1}  Eke, V.R., Lawrence, D.J., Teodoro, L.F.A., How thick are Mercury’s polar water ice deposits?, \emph{Icarus,} \textbf{284}, pp. 407–415, 2017.
\bibitem{ElEff}  Ayres, L.W., Ayres, R.U., Pokrovsky, V., \emph{On the Efficiency of Us Electricity Usage Since 1900,} IR-04-027, 2004.
\bibitem{Truba}  Alibaba Group, \emph{Small diameter welded pipe production line,} 2020,\\
    <https://www.alibaba.com/product-detail/ \\
    Small-diameter-welded-pipe-production-line\_62030527762.html?>\\
    Accessed March 19, 2020.
\bibitem{Tarelka} India Mart, \emph{SS Plate Rolling Machine, Production Capacity: 15-20 Ton/Day,} 2020.\\
    <https://www.indiamart.com/proddetail/ss-plate-rolling-machine-14649122673.html>,\\
    Accessed March 19, 2020.
\bibitem{Orechi} ABM Fasteners, India,  \emph{Nut  Forming Machines (Automatic,  High Speed, Cold Forging),} 2020,\\
    <http://www.abmfasteners.com/Fasteners\_machinery/Nut\_Plant.htm>\\
    Accessed March 23, 2020.
\bibitem{Orechi1} DinStock, Ltd, \emph{Weight Chartfor Hexagon Bolts \& Nuts,} 2020,\\
    <http://www.dinstock.com/useruploads/files/weight\_chart\_for\_hexagon\_bolts\_\&\_nuts.pdf>\\
    Accessed March 23, 2020.
\bibitem{AdMan}  Srivatsan, T.S., Sudarshan, T.S., \emph{Additive Manufacturing: Innovations, Advances, and Applications,} Boca Raton, Florida: CRC Press, 2016.
\bibitem{MAM4}  Gilbert, S.W.,  Thomas, D.S., \emph{Costs and Cost Effectiveness of Additive Manufacturing: A Literature Review and Discussion,} Gaithersburg, MD: U.S. Dept. of Commerce, National Institute of Standards and Technology, 2014.
\bibitem{MAM7}  Sciaky, Inc., \emph{Metal Additive Manufacturing Systems Brochure,}\\
    <https://www.sciaky.com/images/pdfs/product-sheets/Sciaky-EBAM-Systems.pdf>,\\
    Accessed March 21, 2020.
\bibitem{MAM8}  3Dprinting.com, \emph{Desktop Metal System Printer,} \\
    <https://3dprinting.com/products/industrial-3d-printer/desktop-metal-system-printer/>\\
    Accessed March 21, 2020.
\bibitem{MAM9}  x3D Systems, \emph{Metalform750,}\\
    <https://x3d-systems.com/metalform750/>,\\
    Accessed March 21, 2020.
\bibitem{MAMB}  3Dprinting.com, \emph{Concept Laser Mlab Cusing,} \\
    <https://3dprinting.com/products/industrial-3d-printer/concept-laser-mlab-cusing/>\\
    Accessed March 21, 2020.
\bibitem{MAM3}  Andrade-Campos, A., Barroqueiro, B., Neto, V., Valente, R.A.F., Metal Additive Manufacturing Cycle in Aerospace Industry: A Comprehensive Review, Journal of Manufacturing and Materials Processing, \textbf{3}(52), 2019.
\bibitem{MAMC}  Langnau, L., \emph{Additive Manufacturing Handbook 2018,} Design World, 2018.
\bibitem{SLR} Space Launch Report Log by Decade, \emph{Space Launch Report}, 2015.\\
    http://www.spacelaunchreport.com/logdec.html
\bibitem{TotM}Tkatchova, S., \emph{Space-based Technologies and Commercialized Development: Economic Implications and Benefits,} Hershey, PA: Engineering Science Reference, 2011.
\bibitem{LCost}  Jones, H.W.,  The Recent Large Reduction in Space Launch Cost, \emph{48$^{\text{th}}$ International Conference on Environmental Systems,}  8-12 July 2018.
\bibitem{Shuttle} \emph{Gao Report on Analysis of Cost of Space Shuttle Program: Hearings (with Committee's Summary and Conclusions),} Ninety-Third Congress, First Session, Washington: U.S. Govt. Print. Off, 1973.
\bibitem{guide} Fishlock, D., \emph{A Guide to Earth Satellites,} London: Macdonald and Co, 1971.
\bibitem{Cost1} Sadeh, E., \emph{Space Politics and Policy: An Evolutionary Perspective,} Dordrecht: Kluwer Academic Publishers, 2002.
\bibitem{Raport01}\emph{The Annual Compendium of Commercial Space Transportation 2016,} \emph{Federal Aviation Administration,} 2016.\\
    <https://www.faa.gov/about/office\_org/headquarters\_offices/ast/media/2016\_Compendium.pdf>
\bibitem{SLVD}  Woodward, D., \emph{Space Launch Vehicle Design,} Dissertation at Department of Mechanical and Aerospace Engineering University of Texas at Arlington, 2017.
\bibitem{Falcon2} Wall, M., A Sixth Success! SpaceX Again Lands Rocket on a Ship at Sea, \emph{space.com}, August 14, 2016.
\bibitem{Falcon3} Gorn, M.H., Chiara, G.De., \emph{Spacecraft: 100 Iconic Rockets, Shuttles, and Satellites That Put Us in Space,} Minneapolis: Quayside Publishing Group, 2018.
\bibitem{Falcon1} Bignami, G. F. and Sommariva, A., \emph{The Future of Human Space Exploration}, IASF-INAF, Milan, Italy, 2016.
\bibitem{FH01} Wanjek, C., \emph{Spacefarers: How Humans Will Settle the Moon, Mars, and Beyond,} Cambridge, Massachusetts : Harvard University Press, 2020.
\bibitem{FH02} Pratt, T., Allnutt, J.E., \emph{Satellite Communications,}  	Hoboken, NJ : John Wiley \& Sons, Ltd, 2020.
\bibitem{SpaceCannon}  Powell, J., Maise G., Pellegrino, C., \emph{Star Tram: The New Race to Space,} 2013.
\bibitem{NRSL1}  Bolonkin, A., \emph{Non-Rocket Space Launch and Flight,} Elsevier Science, 2010.
\bibitem{RamAc1}  Liu, S., Knowlen, C., \emph{Axis-symmetric Ram Accelerator Projectile Performance Characteristics,} Seattle: University of Washington Libraries, 2019.
\bibitem{RamAc2}  Higgins, A.J., Ram Accelerators: Outstanding Issues and New Directions, \emph{Journal of Propulsion and Power,} \textbf{22}(6), pp. 1170-1187, 2006.
\bibitem{StarTram2010} Maise, G., Powell, J., Rather, J., Maglev Launch: Ultra Low Cost Ultra/High Volume Access to Space for Cargo and Humans, Space, Propulsion, and Energy Sciences International Forum, 2010.
\bibitem{StarTram2010A} del Monte, L., Gamma F., Andriani, R., \emph{Maglev for Space Launchers,} America Institute of Aeronautics \& Astronautics, 2001.
\bibitem{spaceprop} Tajmar, M., \emph{Advanced Space Propulsion Systems.} Wien: Springer, 2003.
\bibitem{LRP} Liquid rocket propellant, \emph{Wikipedia: The Free Encyclopedia.} Wikimedia Foundation, 1 Nov. 2016 \\
    <https://en.wikipedia.org/wiki/Liquid\_rocket\_propellant>
\bibitem{ionhist}  Choueiri, E.Y.,  A Critical History of Electric Propulsion: The First 50 Years (1906-1956), \emph{Journal of Propulsion and Power}, Vol. 20, No. 2, p. 193-203, 2004.
\bibitem{ion1} Lopez, H., \emph{Study of earth-to-mars transfers with low-thrust propulsion,} 2012.\\
    http://upcommons.upc.edu/bitstream/handle/2099.1/14029/Project\_Report.pdf
\bibitem{MPT1} Choueiri, E.Y., New Dawn For Electric Rockets, \emph{Scientific  American} 300.2: p. 58-65, 2009.
\bibitem{MPT2} Andrenucci, M.,   Magnetoplasmadynamic thrusters,  \emph{Encyclopedia of Aerospace Engineering.}\\
    http://www.spaceatdia.org/uploads/mariano/ep2/Magnetoplasmadynamic Thrusters.pdf
\bibitem{Vasya} Variable Specific Impulse Magnetoplasma Rocket, \emph{Wikipedia: The Free Encyclopedia.} Wikimedia Foundation, 1 Nov. 2016 \\
    <https://en.wikipedia.org/wiki/Variable\_Specific\_Impulse\_Magnetoplasma\_Rocket>
\bibitem{Vasya1} Ilin, A.V., Gilman, D.A., Carter, M.D., Chang, F.R., Squire, J.P. and Farrias, J.E., VASIMR Solar Powered Missions for NEA Retrieval and NEA Deflection,  \emph{Presented at the 3rd International Electric Propulsion Conference,} The George Washington University, Washington, D.C., October 6–10, 2013.\\
    <http://www.adastrarocket.com/AndrewIEPC13-336-Paper.pdf>
\bibitem{Next} Patterson,  M.  and  Benson,  S.,  NEXT  Ion  Propulsion  System  Development Status  and  Performance,  AIAA-2007-5199, \emph{43rd AIAA/ASME/SAE/ASEE Joint Propulsion Conference and Exhibit}, 2007. \\
    <https://www.grc.nasa.gov/WWW/ion/pdfdocs/AIAA-2007-5199.pdf>
\bibitem{Vasya2} Squire, J.P.,  Carter, M.D., Chang, F.R., Giambusso, M., Glover, T.W., Ilin, A., Oguilve-Araya, J., Olsen, C.S., Bering, E.A. and Longmier, B.W.  VASIMR Spaceflight Engine System Mass Study and Scaling with Power, \emph{Presented at the 3rd International Electric Propulsion Conference,} The George Washington University, Washington, D.C., October 6–10, 2013.\\
    <http://www.adastrarocket.com/IEPC13-149\_JPSquire\_submit.pdf>
\bibitem{MPT3} Gorshkov, O.A., Kozubsky, K.N., Obukhov, V.A., Ostrovsky, V.G., Shutov, V.N., Development of High Power Magnetoplasmadynamic Thrusters in the USSR, \emph{Presented at the 30th International Electric Propulsion Conference,} Florence, Italy, 2007.
\bibitem{MPDynamic} Nerheim, N.M., Kelly, A.J., \emph{A Critical Review of the Magnetoplasmadynamic (MPD) Thrustor for Space Applications,} Technical Report 32-1196, National Aeronautics And Space Administration, Jet Propulsion Laboratory, California Inititute Of Technology Pasadena, California, 1968.
\bibitem{MPDT02} Mantenieks, M.A., Sovey, J.S., \emph{Performance and Lifetime Assessment of MPD Arc Thruster Technology,} NASA NASA Technical Memorandum 101293 AIAA-88-3211, 1988.
\bibitem{NucSpace01}  White, D.J., Technology Survey and Performance Scaling for the Design of High Power Nuclear Electric Power and Propulsion Systems, Massachusetts Institute of Technology, 2011.
\bibitem{MPDT08} Nada, T.R., Performance characterisation of MPD thrusters, \emph{The Aeronautical Journal,}
    \textbf{111}(1121), pp.443-452, 2007.
\bibitem{ArcJet2} Curran, F.M., Haag, T.W., Hamley, J.A., Sankovic, J.M., Sarmiento, C.J., Hydrogen Arcjet Technology, NASA Technical Memorandum 105340, IEPC-91-018, 1991.
\bibitem{ArcJet4} Auweter-Kurtz, M., Kurtz, H., High Thrust Density Electric Propulsion for Heavy Payload In-Space Transportation, \emph{4$^{\text{th}}$ International Spacecraft Propulsion Conference, Cagliari, Sardinia, Italy}, 2004.
\bibitem{Vasya3} Longmier, B.W., Squire, J.P., Cassady, L.D., Ballenger, M.G., Carter, M.D.,  Olsen, C.,  Ilin, A., Glover, T.W., McCaskill, G.E., Chang, F.R., Bering, E.A. and Del Valle, J., VASIMR VX-200 Performance Measurements and Helicon Throttle Tables Using Argon and Krypton,  \emph{Presented at the 3rd International Electric Propulsion Conference,} The George Washington University, Washington, D.C., October 6–10, 2013.\\
    <http://www.adastrarocket.com/Ben\_IEPC11-156.pdf>
\bibitem{Advanced2020} Bruno, C., Antonio G., A., \emph{Advanced Propulsion Systems and Technologies, Today to 2020,} Reston, Va: American Institute of Aeronautics and Astronautics, 2008.
\bibitem{NTh1}  Borowski, S.K., McCurdy, D.R., Packard, T.D., \emph{Nuclear Thermal Propulsion (NTP): A Proven Growth Technology for Human NEO / Mars Exploration Missions,} 2012 IEEE Aerospace Conference, Big Sky, MT, 2012.
\bibitem{RPA} Ponomarenko, A., \emph{Rocket Propulsion Analysis,} V 1.2.6, Lite Edition 2011,
    <http://www.propulsion-analysis.com/>,
    Cologne, Germany, 2011.
\bibitem{LunCol}  General Dynamics Corporation, Lyndon B. Johnson Space Center, \emph{Lunar Resources Utilization for Space Construction: Final Presentation, 21 February 1979,} San Diego, Calif: The Division, 1979.
\bibitem{LunCol1}  Dalton, C., Hohmann, E., \emph{Conceptual Design of a Lunar Colony,} NASA/ASEE Systems Design Institute, 1972.
\bibitem{LunCol2}  Eckart, P., Eckart, P., Aldrin, B., \emph{The Lunar Base Handbook: An Introduction to Lunar Base Design, Development, and Operations,} New York: McGraw-Hill, 1999.
\bibitem{LunCol3}  Klepeis, A., \emph{Moon Base and Beyond: The Lunar Gateway to Deep Space,} North Mankato, MN : Capstone Press, 2020.
\bibitem{LunarIce}  Crider, D.H., Vondrak, R.R., Ice at the Lunar Poles, \emph{American Scientist,} \textbf{91}, p.322-329, 2003.
\bibitem{LunaResources1}  Fishman, C., \emph{The Big Thirst: The Secret Life and Turbulent Future of Water,} Free Press, 2014.
%\bibitem{LunarH}  Greenhagen, B.T., Lawrence, D.J., Maurice, S., Peplowski, P.N., Prettyman, T.H., Bulk hydrogen abundances in the lunar highlands: Measurements from orbital neutron data, \emph{Icarus}, \textbf{255}, pp.127-134, 2015.
%\bibitem{LunarBase1}  Billingham, J., Gilbreath, W., Gossett, B., O'Leary, B., Eds., \emph{Space Resources and Space Settlements,} National Aeronautics and Space Administration Scientific and Technical Information Branch Washington, D.C., 1979.
\bibitem{LunarAu} Brandstatter, F., Demidova, S.I., Kurat, G., Lorenz, C.A., Nazarov, M.A., Ntaflos, Th., Chemical Composition of Lunar Meteorites and the Lunar Crust, \emph{Petrologiya,} \textbf{15}(4), p.416–437, 2007.
%\bibitem{HeatPipe1}  Reay, D., \emph{Heat Pipes,} Oxford: Butterworth-Heinemann, LTD,  2017.
%\bibitem{MoonGlass}  Bennett, N., Buckman, J., Cowley, A., Gibbons, J., Groetsch, A., Schleppi, J.,  Manufacture of Glass and Mirrors from Lunar Regolith Simulant, \emph{Journal of Materials Science,} \textbf{54}(5) pp.3726-3747, 2019.
\bibitem{MoonGlass1} Landis, G.A., \emph{Materials Refining for Solar Array Production on the Moon,} National Aeronautics and Space Administration: NASA/TM—2005-214014, 2005.
\bibitem{Renogy} Renogy, Inc., 2775 E. Philadelphia St, Ontario, CA 91761.\\
    <http://www.renogy.com>, Accessed March 3, 2020.
\bibitem{LwCell1}  Edmondson, K.M., Glenn, G., Haddad, M.H., Isshiki, T.D., Karam, N.H., King, R.R., Law, D.C., Siddiqi, N., \emph{Lightweight, Flexible, High-Efficiency III-V multijunction cells,}
    Proceedings of 4th World Conference on Photovoltaic Energy Conversion, Waikoloa, Hawaii, May 7-12,  pp.1879-1882, 2006.
\bibitem{LwCell2}    Barnes, T.M., Booth, S., Dabney, M.S., Feldman, D., Glynn, S., Haegel, N.M., Kempe, M.D., McGott, D.L., Reese, M.O., Increasing Markets and Decreasing Package Weight for High Specific Power Photovoltaics, \emph{Nature Energy,} \textbf{3}(11), pp.1002-1012, 2018.
\bibitem{LunarPV3}  Landis, G.A., Perino, M.A., \emph{Lunar Production of Solar Cells,} National Aeronautics and Space Administration, Report NASA TM-102102, 1989.
\bibitem{LunarPV4}  Duke, M., Freundlich, A., Ignatiev, A., Rosenberg, S., \emph{New Architecture for Space Solar Power Systems: Fabrication of Silicon Solar Cells Using In-Situ Resources,} NIAC 2nd Annual Meeting, June 6-7 2000.
\bibitem{LunarPV1}  Duke, M., Ignatiev, A., Rosenberg, S.,  \emph{In-Situ Electric Power Generation To Support Solar System Exploration and Colonization: Manufacture of Thin Film Silicon Solar Cells On The Moon,}  ICEUM4, ESTEC, Noordwijk, The Netherlands, 10-15 July 2000.
\bibitem{HPonic}  Ming, D. W., Henninger, D. L., \emph{Lunar Base Agriculture: Soils for Plant Growth,} Madison, Wis., USA: American Society of Agronomy, 1989.
\bibitem{SFict1}  Wolfe, G.K., \emph{How Great Science Fiction Works,} The Great Courses, Chantilly, VA, 2016.
\bibitem{SFict2}  Brake, M., \emph{The Science of Science Fiction: The Influence of Film and Fiction on the Science and Culture of Our Times,} Skyhorse Publishing, New York, NY, 2018.


%\bibitem{ShubovE} Shubov, M., History of Prime Movers and Future Implications, \emph{International Journal of Recent Scientific Research,} \textbf{11}(01 B), pp. 36751-36759, 2020.
%\bibitem{MercIce3}  Chabot, N. L., Shread, E. E., Harmon, J. K., Investigating mercury's south polar deposits: arecibo radar observations and high-resolution determination of illumination conditions, \emph{Journal of Geophysical Research, Planets}, \textbf{123}(2), pp. 666-681, 2018.
%\bibitem{MercIce2}  NASA, MESSENGER finds new evidence for water ice, organic material at Mercury's poles, \emph{Phys.org,} 2012.\\
%    <https://phys.org/news/2012-11-messenger-evidence-ice-mercury-poles.html>\\
%    Accessed Feb 10, 2020.
%\bibitem{MnSteels}  Kirillov, Y.V., Apaev, B.A., Madyanov, S.A., Effect of Austenitizing on Softening of Chromium and Mangenese Steels During Tempering, \emph{Metal Science and Heat Treatment,} \textbf{20}(7), pp.576-578, 1978.
%\bibitem{CorrNa1}  \emph{High-temperature Characteristics of Stainless Steels,} Washington, DC: American Iron and Steel Institute, 1992.
%\bibitem{CorrNa2}  Zimmerman, C.A., \emph{Corrosion of Type 316 Stainless Steel in Nak Service: A Literature Survey,} Idaho Falls, Idaho: U.S. Atomic Energy Commission, Idaho Operations Office, 1965.
%\bibitem{BlastF}  Surendranathan, A. O., \emph{An Introduction to Ceramics and Refractories,} Boca Raton: CRC Press, Taylor et Francis, 2015.
%\bibitem{Roasting}  Marsden, J., House, I., \emph{The Chemistry of Gold Extraction,} Littleton, Colo: Society for Mining, Metallurgy, and Exploration, 2006.
%\bibitem{GrAnode} Pierson, H.O., \emph{Handbook of Carbon, Graphite, Diamond, and Fullerenes: Properties, Processing, and Applications,} Park Ridge, N.J: Noyes Publications, 1993.
%\bibitem{COComb} Saastamoinen, J.J., New Simplified Rate Equation for Gas-Phase CO Oxidation at Combustion, \emph{Energy \& Fuels,} \textbf{14}, pp. 1156-1160, 2000.
%\bibitem{CORocket} Linne, D.L., \emph{Carbon Monoxide and Oxygen Combustion Experiments: A Demonstration of Mars in Situ Propellants,} Washington, D.C.: National Aeronautics and Space Administration, 1991.
%\bibitem{Exergym} Szargut J., International progress in second law analysis, \emph{\emph{Energy,}} Vol. 5, No. 8-9, pp. 709-718, 1980.
%\bibitem{Exergy} Gundersen, T., \emph{An Introduction to the Concept of Exergy and Energy Quality,} Department of Energy and Process Engineering Norwegian University of Science and Technology, Trondheim, Norway, 2009.
%\bibitem{HLiq07}  Krasae-in, S., Neksa, P., Stang, J.H., "Development of large-scale hydrogen liquefaction processes from 1898 to 2009," \emph{International Journal of Hydrogen Energy, }·Vol. 35, 2010.
%\bibitem{UF4-01} Anghaie, S.: Development of Liquid-Vapor Core Reactors with MHD Generator for Space Power and Propulsion Applications, \emph{Final report of DOE project DE-FG07-98ID 13635}, 2002.\\
%    <http://www.osti.gov/scitech/servlets/purl/799231>
%\bibitem{VacIns01}  Augustynowicz, S.D.,  Fesmire, J.E., "Cryogenic Thermal Insulation Systems,"  16$^{th}$ Thermal and Fluids Analysis Workshop, Orlando, Florida, 2005.
%\bibitem{VacIns06}  Augustynowicz, S. D., Fesmire, J. E., Cryogenic Insulation System for Soft Vacuum, \emph{Advances in Cryogenic Engineering,} \textbf{45}, pp. 1691-1698, 2000.
%\bibitem{VacIns02}  Fesmire, J.E., "Layered Thermal Insulation Systems for Industrial and Commercial Applications," NASA Kennedy Space Center, 2017.
%\bibitem{VacIns05}  Fesmire, J.E., "Standardization in cryogenic insulation systems testing and performance data," 25th International Cryogenic Engineering Conference and the International Cryogenic Materials Conference, 2014.
%\bibitem{VacIns07}  Fesmire, J.E., Standardization in Cryogenic Insulation Systems Testing and Performance Data, \emph{Physics Procedia,} \textbf{67}, pp. 1089-1097, 2015.
%\bibitem{VacIns04}  \emph{Integrated Insulation System for Cryogenic Automotive Tanks (iCAT),} Report DE-EE0007649,  Vencore Services and Solutions, Inc., 2018.
%\bibitem{Venus1}  Sharma, S., Ramanan, R.V., Venus Gravity Assist Transfers to Mercury Using Differential Evolution, \emph{First International Conference on Recent Advances in Aerospace Engineering (ICRAAE)}, pp. 135-139, 2017.
%\bibitem{EarthMerc}  Williams, D.R., Mercury Fact Sheet, \emph{NASA,} 2018.\\ <https://nssdc.gsfc.nasa.gov/planetary/factsheet/mercuryfact.html>,\\
%    Accessed Februry 25, 2020.
%\bibitem{hoh}Hohmann, W., \emph{The Attainability of Heavenly Bodies,} Washington, D.C: National Aeronautics and Space Administration, 1960.
%\bibitem{Venus2}  National Aeronautics and Space Administration, Scientific and Technical Information Division, \emph{Planetary Flight Handbook, Pt. 9, Direct and Venus Swingby Trajectories to Mercury,} Washington: NASA, 1970.
%\bibitem{Venus3}  Biesbroek, R., \emph{Lunar and Interplanetary Trajectories,} Springer International Publishing, Cham, Switzerland, 2016.
%\bibitem{NuPr}   Lawrence, T.J., \emph{Nuclear Thermal Rocket Propulsion Systems,} U.S. Air Force Academy Department of Astronautics, USAFA, Colorado, 2005.
\end{thebibliography}
\end{document}